\documentclass{article}

\usepackage[preprint]{neurips_2025}

\usepackage{natbib}
\usepackage[utf8]{inputenc} 
\usepackage[T1]{fontenc}    
\usepackage{hyperref}       
\usepackage{url}            
\usepackage{booktabs}       
\usepackage{amsfonts}       
\usepackage{nicefrac}       
\usepackage{microtype}      
\usepackage[table]{xcolor}         
\usepackage{multirow}
\usepackage{pgfplots}
\usepackage{subcaption}
\usepackage[most]{tcolorbox}
\usepackage{makecell}
\usepackage{graphicx}

\newcommand{\recability}{\textsc{Recability}}
\newcommand{\recabilities}{\textsc{Recabilities}}
\newcommand{\recbench}{\textsc{RecBench}}
\newcommand{\recbenchmd}{\textsc{RecBench-MD}}
\newcommand{\llmasrs}{\textsc{LLM-as-RS}}
\newcommand{\llmforrs}{\textsc{LLM-for-RS}}

\newcommand{\bertbase}{BERT$_\text{base}$}
\newcommand{\optbase}{OPT$_\text{350M}$}
\newcommand{\optlarge}{OPT$_\text{1B}$}
\newcommand{\llamayi}{Llama-1$_\text{7B}$}
\newcommand{\llamaer}{Llama-2$_\text{7B}$}
\newcommand{\llamasan}{Llama-3$_\text{8B}$}
\newcommand{\llamasy}{Llama-3.1$_\text{8B}$}
\newcommand{\qwensmall}{Qwen-2$_\text{500M}$}
\newcommand{\qwenbase}{Qwen-2$_\text{1.5B}$}
\newcommand{\qwen}{Qwen-2$_\text{7B}$}
\newcommand{\glm}{GLM-4$_\text{9B}$}
\newcommand{\mistral}{Misrtal-2$_\text{7B}$}
\newcommand{\chatgpt}{GPT-3.5}
\newcommand{\dsqwen}{DS-Qwen-2$_\text{7B}$}
\newcommand{\recgpt}{RecGPT$_\text{7B}$}
\newcommand{\pwu}{P5$_\text{beauty}$}
\newcommand{\recformer}{Recformer}
\newcommand{\ewu}{E5$_\text{base-v2}$}
\newcommand{\phier}{Phi-2$_\text{3B}$}

\newcommand{\domain}[1]{\raisebox{-0.5ex}{\includegraphics[height=1.2em]{Assets/Images/Domain/#1.pdf}}}
\newcommand{\dNews}{\domain{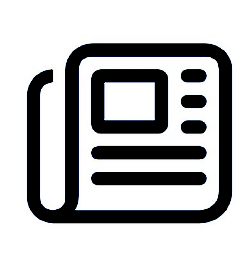}}
\newcommand{\dBook}{\domain{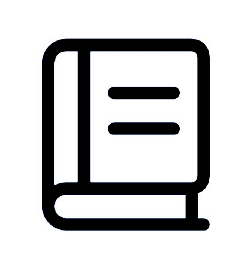}}
\newcommand{\dFashion}{\domain{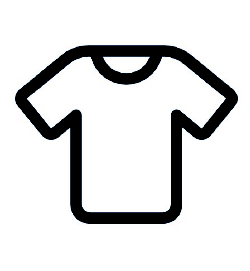}}
\newcommand{\dHotel}{\domain{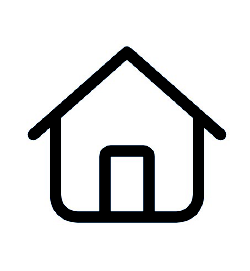}}
\newcommand{\dMovie}{\domain{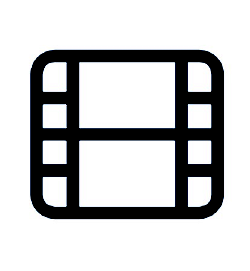}}
\newcommand{\dVideo}{\domain{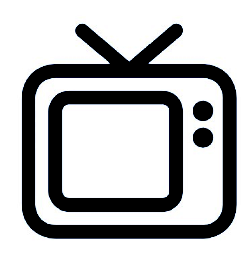}}
\newcommand{\dFood}{\domain{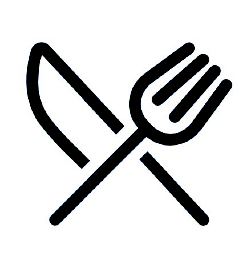}}
\newcommand{\dGame}{\domain{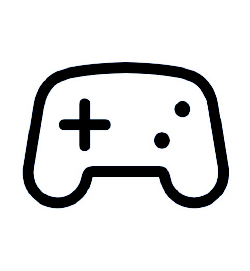}}
\newcommand{\dElec}{\domain{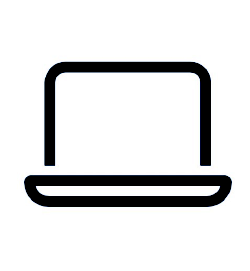}}
\newcommand{\dMusic}{\domain{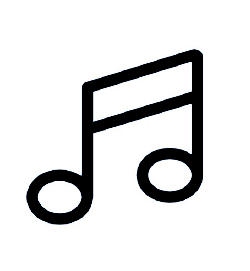}}

\newcommand{\mind}{\dNews{} MIND}
\newcommand{\microlens}{\dVideo{} Micro.}
\newcommand{\goodreads}{\dBook{} Good.}
\newcommand{\cds}{\dMusic{} CDs}
\newcommand{\hm}{\dFashion{} H\&M}
\newcommand{\movielens}{\dMovie{} Movie.}
\newcommand{\yelp}{\dFood{} Yelp}
\newcommand{\steam}{\dGame{} Steam}
\newcommand{\electronics}{\dElec{} Elec.}
\newcommand{\hotelrec}{\dHotel{} Hotel.}
\newcommand{\pens}{\dNews{} PENS}
\newcommand{\netflix}{\dVideo{} Netflix}
\newcommand{\books}{\dBook{} Books}
\newcommand{\lastfm}{\dMusic{} Last.fm}
\newcommand{\pog}{\dFashion{} POG}

\definecolor{kbgA}{RGB}{240, 220, 220}  
\definecolor{kbgB}{RGB}{250, 235, 200}  
\definecolor{kbgC}{RGB}{230, 240, 205}  
\definecolor{kbgD}{RGB}{210, 235, 230}  
\definecolor{kbgE}{RGB}{215, 230, 245}  
\definecolor{kbgF}{RGB}{230, 220, 245}  
\definecolor{kbgG}{RGB}{240, 240, 240}  
\definecolor{kbgH}{RGB}{245, 225, 235}  

\definecolor{bgGood}{RGB}{160, 190, 135}
\definecolor{bgIgnore}{RGB}{180, 165, 160}

\definecolor{firstordercolor}{RGB}{96,125,139}  
\definecolor{secondordercolor}{RGB}{183,110,121}  
\definecolor{barcolor}{RGB}{69,90,100}

\definecolor{bglight}{RGB}{248, 248, 248}

\newcommand{\regime}[1]{\raisebox{-0.5ex}{\includegraphics[height=1em]{Assets/Images/Regime/#1.pdf}}}

\newcommand{\rA}{\regime{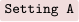}}
\newcommand{\rB}{\regime{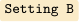}}
\newcommand{\rC}{\regime{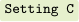}}
\newcommand{\rD}{\regime{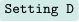}}
\newcommand{\rE}{\regime{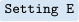}}
\newcommand{\rF}{\regime{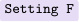}}
\newcommand{\rG}{\regime{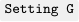}}
\newcommand{\rH}{\regime{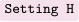}}

\title{Evaluating \recabilities{} of Foundation Models: \\ A Multi-Domain, Multi-Dataset Benchmark}

\author{
Qijiong Liu\textsuperscript{1},
Jieming Zhu\textsuperscript{2},
Yingxin Lai\textsuperscript{3},
Xiaoyu Dong\textsuperscript{1},\\
\textbf{Lu Fan}\textsuperscript{1},
\textbf{Zhipeng Bian}\textsuperscript{4},
\textbf{Zhenhua Dong}\textsuperscript{2},
\textbf{Xiao-Ming Wu}\textsuperscript{1}\\
\textsuperscript{1}The Hong Kong Polytechnic University
\textsuperscript{2}Huawei Noah's Ark Lab, Shenzhen, China \\ \textsuperscript{3}Xiamen University, Xiamen, China 
\textsuperscript{4}Shenzhen University, Shenzhen, China \\
\texttt{liu@qijiong.work}
}

\begin{document}


\maketitle

\begin{abstract}

Comprehensive evaluation of the recommendation capabilities of existing foundation models across diverse datasets and domains is essential for advancing the development of recommendation foundation models. In this study, we introduce \recbenchmd{}, a novel and comprehensive benchmark designed to assess the recommendation abilities of foundation models from a zero-resource, multi-dataset, and multi-domain perspective. Through extensive evaluations of 19 foundation models across 15 datasets spanning 10 diverse domains—including e-commerce, entertainment, and social media—we identify key characteristics of these models in recommendation tasks. Our findings suggest that in-domain fine-tuning achieves optimal performance, while cross-dataset transfer learning provides effective practical support for new recommendation scenarios. Additionally, we observe that multi-domain training significantly enhances the adaptability of foundation models.
All code\footnote{\url{https://github.com/Jyonn/RecBench-MD}} and data\footnote{\url{https://www.kaggle.com/datasets/qijiong/recbench-md}} have been publicly released to facilitate future research.

\end{abstract}







\section{Introduction}

The rapid emergence of foundation models, particularly large language models (LLMs), has revolutionized various fields such as natural language processing (NLP)~\citep{llm-llama-1,reid2024gemini1_5} and computer vision~\citep{kirillov2023segment,li2024llava}.
Recently, their application in recommender systems has attracted considerable interest, as these models promise a unified framework capable of modeling user--item interactions through natural language~\citep{survey1,survey2,survey3}.
Despite the existence of numerous foundation models, most are primarily designed for NLP tasks, and there is currently a lack of effective strategies for selecting appropriate models to develop recommendation foundation models. Consequently, assessing the recommendation abilities, referred to as \recabilities{}, of foundation models has become increasingly important.

Recommendation foundation models, akin to LLMs with general NLP capabilities, should exhibit broad \textbf{zero-resource} \footnote{In this paper, \textit{zero-resource} means fine-tuning on some datasets and testing on unseen ones (cross-dataset), while \textit{zero-shot} means testing without any fine-tuning.} \recabilities{}, allowing for inference on unseen datasets or even novel domains. This necessitates a comprehensive evaluation of the recommendation abilities of existing foundation models across various datasets, domains, training and evaluation strategies, and recommendation tasks and approaches. Although efforts such as LLMRec \citep{bm-llmrec} and PromptRec \citep{bm-promptrec} exist, these studies primarily concentrate on a single domain or dataset using one recommendation approach, leading to a constrained evaluation scope and partial conclusions.



\begin{figure*}
   \centering
   \includegraphics[width=\linewidth]{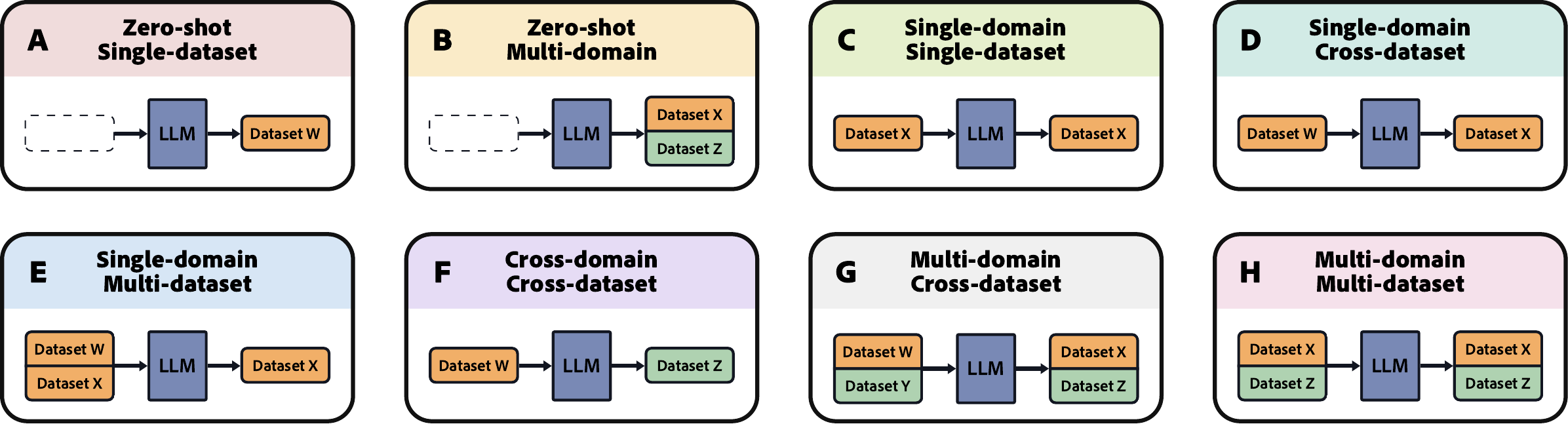}
   \caption{Illustrations of diverse recommendation settings, with colors denoting different domains.}
   \label{fig:regimes}
\end{figure*}

To address these challenges, we introduce a multi-domain recommendation taxonomy that examines all applicable scenarios across eight settings, as depicted in Figure~\ref{fig:regimes}, ranging from \rA{} to \rH{}. Initially, recommendation models were developed for individual datasets, corresponding to \rC{} (single-domain single-dataset). Subsequently, researchers explored cross-domain recommendation models, represented by \rF{} (cross-domain cross-dataset), which aim to transfer user interest knowledge from a source domain to a target domain. Additionally, some studies have investigated the integration of multiple domains to train a unified model, corresponding to \rH{} (multi-domain multi-dataset), transitioning from a one-dataset-one-model paradigm to a multiple-dataset-one-model paradigm. More recently, several studies have assessed the zero-shot \recabilities{} of foundation models. However, these evaluations are often limited to a single dataset (\rA{}), which can lead to unreliable and biased results. A more robust approach is to evaluate across multiple datasets and compute the average, as illustrated in \rB{}, i.e., zero-shot multi-domain.


In this work, we present a comprehensive benchmark, \recbenchmd{}, specifically designed to evaluate the \recabilities{} of foundation models from a zero-resource, multi-dataset, and multi-domain perspective, encompassing all settings illustrated in Figure~\ref{fig:regimes}. This study is pioneering in its benchmarking of cross-dataset recommendation for zero-resource settings. We have specifically examined a range of recommendation approaches, including prompt-based ranking tasks and embedding-based matching tasks, thereby covering the main recommendation scenarios. Our evaluation is unprecedented in scope, encompassing 15 recommendation datasets across 10 domains and 19 foundation models. Furthermore, we provide open-source code and datasets, facilitating easy evaluation for future large-scale recommendation or foundation models with simple configuration. Our experiments required an impressive 1,000 GPU hours, and the platform’s reusability significantly reduces experimental costs for future researchers, allowing them to concentrate more on model optimization and algorithm innovation.


Our benchmarking results reveals several key insights.
\textbf{First,} larger models tend to benefit more from joint training on multiple datasets or domains, exhibiting stronger cross-domain generalization.
\textbf{Second,} the degree of transferability across domains varies considerably, with a strong dependence on the characteristics of the source dataset.
\textbf{Third,} while in-domain datasets exhibit higher relevance, this is not universally observed across all scenarios.
\textbf{Fourth,} cross-dataset transfer can serve as an effective model warm-up strategy in novel recommendation contexts, though it is challenging to exceed the performance upper bound established by fine-tuning on single or multiple datasets within the target domain.


\section{Related Work}

\textbf{Existing Benchmarks.} Several benchmarks have been proposed to evaluate the \recabilities{} for foundation models, including LLMRec~\citep{bm-llmrec}, PromptRec~\citep{bm-promptrec}, and others~\citep{bm-lmasrs,bm-llmasrs,bm-rsbench}. However, as illustrated in Table~\ref{tab:benchmark-list}, these benchmarks \textbf{i)} provide only a limited evaluation of recommendation settings, often focusing on a single approach. In addition, \textbf{ii)} the number of foundation models and datasets evaluated remains relatively small, resulting in an incomplete and fragmented performance landscape in this domain.

\textbf{Multi-domain Recommendation.}
Traditional multi-domain recommendation methods predominantly rely on item-based or user-based knowledge transfer, using common items or shared user interactions to mitigate data sparsity and domain discrepancies~\citep{cdr-user1,cdr-user2,cdr-user3}.
However, such approaches require explicit entity-level overlap between domains--a condition rarely met in real-world scenarios~\citep{surveycdr1,surveycdr2}.
In contrast, text-based knowledge transfer leverages rich entity-side information, such as item descriptions and user profiles, used in diverse features~\citep{cdr-non1,cdr-non2}.
Borrowing the semantic comprehension and generation capabilities of foundation models, text-based methods boost cross-domain learning without the need for explicit entity alignment, i.e., non-overlap for users and items, thereby offering a more flexible and robust framework for transferring knowledge across heterogeneous domains.

\definecolor{checkmark}{RGB}{34, 139, 34}
\definecolor{cross}{RGB}{220, 20, 60}
\definecolor{explain}{RGB}{184, 134, 11} 

\newcommand{\support}{\textcolor{checkmark}{$\checkmark$}}
\newcommand{\nosupport}{\textcolor{cross}{$\times$}}
\newcommand{\explain}{\textcolor{explain}{--}}
\newcommand{\bmgood}[1]{\textcolor{checkmark}{#1}}
\newcommand{\bmbad}[1]{\textcolor{cross}{#1}}

\begin{table*}[t]
\centering
\renewcommand{\arraystretch}{1.2} 
\setlength\tabcolsep{3pt}

\caption{Comparison of our \recbenchmd{} with existing benchmarks. ``\explain'' indicates that RecBole-CDR theoretically supports the corresponding feature, although no experimental results are provided.
}
\label{tab:benchmark-list}

\resizebox{\linewidth}{!}{
\begin{tabular}{llccccccccc}
\toprule
\textbf{Benchmark} &  & \citeauthor{bm-lmasrs} & OpenP5 & {LLMRec} & {PromptRec} & {\citeauthor{bm-llmasrs}} & {RSBench} & RecBole-CDR & \recbench{} & \recbenchmd{} \\
\textbf{Year} & & \citeyear{bm-lmasrs} & \citeyear{openp5} & \citeyear{bm-llmrec} & \citeyear{bm-promptrec} & \citeyear{bm-llmasrs} & \citeyear{bm-rsbench} & \citeyear{bm-recbole} & \citeyear{bm-recbench} & \textit{(ours)} \\
\midrule
\multirow{2}{*}{\textbf{Scale}} 
 & \#Foundation Models & 4 & 2 & 7 & 4 & 7 & \bmbad{1} & \bmbad{0} & \bmgood{\textbf{17}} & \bmgood{\textbf{19}} \\
 & \#Dataset & \bmbad{1} & 10 & \bmbad{1} & 3 & 4 & 3 & 3 & 5 & \bmgood{\textbf{15}} \\
\midrule
\multirow{6}{*}{\textbf{Setting}} 
 & Zero-shot & \support{} & \nosupport{} & \support{} & \support{} & \nosupport{} & \nosupport{} & \nosupport{} & \support{} & \support{} \\
 & Single-Dataset & \support{} & \support{} & \support{} & \nosupport{} & \support{} & \support{} & \explain{} & \support{} & \support{} \\
 & In-domain Cross-dataset & \nosupport{} & \nosupport{} & \nosupport{} & \nosupport{} & \nosupport{} & \nosupport{} & \explain{} & \nosupport{} & \support{} \\
 & In-domain Multi-dataset & \nosupport{} & \nosupport{} & \nosupport{} & \nosupport{} & \nosupport{} & \nosupport{} & \explain{} & \nosupport{} & \support{} \\
 & Cross-domain & \nosupport{} & \nosupport{} & \nosupport{} & \support{} & \nosupport{} & \nosupport{} & \explain{} & \nosupport{} & \support{} \\
 & Multi-domain & \nosupport{} & \support{} & \nosupport{} & \nosupport{} & \nosupport{} & \nosupport{} & \explain{} & \nosupport{} & \support{} \\
\midrule
\multirow{2}{*}{\textbf{Approach}}
 & Prompt-based & \support{} & \support{} & \support{} & \support{} & \support{} & \support{} & \nosupport{} & \support{} & \support{} \\
 & Embedding-based & \nosupport{} & \nosupport{} & \nosupport{} & \nosupport{} & \nosupport{} & \nosupport{} & \explain{} & \nosupport{} & \support{} \\
\midrule
\multirow{2}{*}{\textbf{Metric}} & Quality & \support{} & \support{} & \support{} & \support{} & \support{} & \support{} & \explain{} & \support{} & \support{} \\
 & Efficiency & \nosupport{} & \nosupport{} & \nosupport{} & \nosupport{} & \nosupport{} & \nosupport{} & \nosupport{} & \support{} & \support{} \\
\bottomrule
\end{tabular}
}
\end{table*}

\textbf{Foundation Models for Recommendation.}
In recent years, integrating large language models (LLMs) into recommender systems has attracted significant academic and industrial interest.
These integrations can be broadly classified into two paradigms~\citep{survey1,survey2,survey3,survey4}: \llmforrs{} and \llmasrs{}. The \llmforrs{} paradigm enhances traditional recommenders via feature engineering or encoding techniques using LLMs~\citep{llmrec,once,store,legommenders,wu2023leveraging,recbase,hu2024lightweight}. In contrast, the \llmasrs{} paradigm employs LLMs directly as recommenders~\citep{llm-recgpt,recformer,p5,mm4rec}. Studies have demonstrated its superior accuracy in contexts such as cold-start scenarios~\citep{tallrec} and in tasks requiring natural language understanding and generation~\citep{luo2023unlocking,wang2023rethinking,he2023large}.

\section{Proposed Benchmark: \recbenchmd{}}

\subsection{Recommendation Settings}

In bottom-level text-based knowledge transfer, we can freely collect training data as long as: (i) each item is described by textual content, and (ii) each user is represented by their item consumption sequence. To systematically explore how cross-domain data influences target-domain recommendation performance, we propose a novel taxonomy comprising eight fine-tuning settings, as illustrated in Figure~\ref{fig:regimes}:

\noindent\textbf{\rA{} (Zero-resource) Zero-shot Single-dataset.} The model is directly evaluated on a single dataset without any fine-tuning. This setting measures the model’s intrinsic \recability{}.

\noindent\textbf{\rB{} (Zero-resource) Zero-shot Multi-domain.} A more comprehensive zero-shot evaluation: the model is tested on multiple datasets from different domains, and performance is averaged to assess generalization.

\noindent\textbf{\rC{} Single-domain Single-dataset.} Fine-tuning and evaluation are performed on the same dataset. This setting reflects standard in-domain supervised learning.

\noindent\textbf{\rD{} (Zero-resource) Single-domain Cross-dataset.} The model is fine-tuned on one or more datasets within a domain and evaluated on a different dataset from the same domain. It assesses domain-level generalization across datasets.

\noindent\textbf{\rE{} Single-domain Multi-dataset.} Training and testing data are drawn from multiple datasets within the same domain, with potential overlap. This setting measures the benefit of aggregating in-domain data.

\noindent\textbf{\rF{} (Zero-resource) Cross-domain Cross-dataset.} The model is fine-tuned on one domain and tested on a completely different one. This setting probes cross-domain transferability of recommendation knowledge.

\noindent\textbf{\rG{} (Zero-resource) Multi-domain Cross-dataset.} Training and testing datasets come from overlapping but non-identical domains. This setting evaluates how auxiliary domain knowledge contributes to target performance when datasets do not overlap.

\noindent\textbf{\rH{} Multi-domain Multi-dataset.} Both domains and datasets overlap between training and testing. This setting examines the upper-bound performance achievable via comprehensive domain and dataset fusion.

\subsection{Recommendation Approaches}

We evaluate \recabilities{} of foundation models with the pair-wise user--item click prediction task. It involves
the estimation of the probability $\hat{y}$ that a user will interact positively with a candidate item. Therefore, the models will be trained by the binary cross-entropy (BCE) loss, formulated as:
\begin{equation}
    \mathcal{L} = -\sum \left[ y \log \hat{y} + (1 - y) \log (1 - \hat{y}) \right],
\end{equation}
where $y \in \{0, 1\}$ denotes the ground-truth label. Borrowing the idea from conventional recommendation, including matching-based and ranking-based models, we devise two recommendation approaches to calculate the click probabilities.

\textbf{Prompt-based Recommendation.}
We concatenate the user sequence with the candidate item where each item in the sequence and the candidate item are represented by their textual feature. Then, the entire user-item sequence will be in conjunction with a task-specific instruction (e.g., \textit{"Will the user be interested in this item? Answer (Yes or No):"}). Next, the model is guided to predict specific output tokens (i.e., ``Yes'' or ``No''), and their corresponding logits, $l_{\text{yes}}$ and $l_{\text{no}}$. Finally, the click probability $\hat{y}$ can be denoted as:
\begin{equation}
\hat{y} = \frac{e^{l_{\text{yes}}}}{e^{l_{\text{yes}}} + e^{l_{\text{no}}}}.
\end{equation}

\textbf{Embedding-based Recommendation.}
Following matching-based two-tower paradigm, here the foundation models are employed as user and item encoders learn their dense representations (embeddings) within a shared latent space. Specifically, we use the last token output embedding for user/item representation when the input is the user sequence or the candidate item. The click probability can be subsequently measured by the cosine similarity:
\begin{equation}
\hat{y} = \frac{\mathbf{u} \cdot \mathbf{t}}{\|\mathbf{u}\| \|\mathbf{t}\|},
\end{equation}
where $\cdot$ denotes the dot product operation, $\|\|$ represents the L2 norm, and $\mathbf{u}$ and $\mathbf{t}$ are user and item representations.

\begin{table*}[t]
\centering
\renewcommand{\arraystretch}{1.2} 

\caption{Datasets evaluated or finetuned in our benchmark.}\label{tab:dataset-list}

\resizebox{\linewidth}{!}{
\begin{tabular}{llllllllll}
\toprule
\multirow{2}{*}{\textbf{Dataset}} & \multirow{2}{*}{\textbf{Domain}} & \multirow{2}{*}{\textbf{Symbol}} & \multicolumn{3}{l}{\textbf{Test set}} & \multicolumn{3}{l}{\textbf{Finetune set}} & \multirow{2}{*}{\textbf{Used Attributes}} \\
\cmidrule(lr){4-6} \cmidrule(lr){7-9}
 &  &  & \#Sample & \#Item & \#User & \#Sample & \#Item & \#User & \\
\midrule
H\&M & Fashion & \hm{} & 20,000 & 15,305 & 5,000 & 100,000 & 50,319 & 25,000 & \texttt{detail\_desc} \\
MIND & News & \mind{} & 20,006 & 3,088 & 1,514 & 100,000 & 5,481 & 7,606 & \texttt{title} \\
MicroLens & Video & \microlens{} & 20,000 & 11,073 & 5,000 & 100,000 & 18,658 & 25,000 & \texttt{title} \\
Goodreads & Book & \goodreads{} & 20,009 & 12,984 & 1,736 & 100,005 & 40,322 & 8,604 &  \texttt{original\_title} \\
Amazon CDs & Music & \cds{} & 20,003 & 15,568 & 4,930 & 100,003 & 55,428 & 24,618 & \texttt{title} \\
\midrule
MovieLens & Movie & \movielens{} & 20,008 & 4,300 & 2,251 & -  & - & - & \texttt{title} \\
Yelp & Restaurant & \yelp{} & 20,003 & 15,239 & 4,013 & - & - & - &  \texttt{name} \\
Steam & Game & \steam{} & 20,000 & 2,216 & 5,000 & - & - & - & \texttt{game\_name} \\
Amazon Electronics & E-commerce & \electronics{} & 20,002 & 11,045 & 5,431 & - & - & - &  \texttt{title} \\
HotelRec & Hotel & \hotelrec{} & 20,002 & 17,295 & 5,437 & - & - & - &  \texttt{name, location} \\
\midrule
POG & Fashion & \pog{} & - & - & - & 100,002 & 15,846 & 15,734 & \texttt{title\_en} \\
PENS & News & \pens{} & - & - & - & 100,007 & 9,053 & 8,542 & \texttt{title} \\
Netflix & Video & \netflix{} & - & - & - & 100,010 & 3,645 & 13,424 & \texttt{title} \\
Amazon Books & Book & \books{} & - & - & - & 100,002 & 28,471 & 25,139 & \texttt{title} \\
LastFM & Music & \lastfm{} & - & - & - & 100,100 & 94,319 & 910 & \texttt{track, artist} \\
\bottomrule
\end{tabular}
}
\end{table*}

\section{Experimental Setup} \label{sec:exp}

\textbf{Datasets.}
 To meaningfully probe foundation model capabilities in recommendation beyond prevalent single-dataset evaluations, a deliberately heterogeneous suite of 15 public datasets across 10 domains was assembled. This collection's scale and diversity (Table~\ref{tab:dataset-list}) are necessary to stress-test the central premise of foundation model generalization across varied recommendation contexts, spanning high-volume consumer arenas (e.g., fashion, news) to specialized niches (e.g., games, hotels). The inherent heterogeneity manifesting in item taxonomies, user interaction dynamics, textual signal richness, and sparsity levels is instrumental, leveraged to transcend potentially idiosyncratic single-domain observations and evaluate genuine cross-domain. For each dataset, the fine-tuning set is randomly split into a training set and validation set in a 9:1 ratio.

\textbf{Foundation Models.}
We collected 19 foundation models from different perspectives to evaluate their \recabilities{}, including: \bertbase{}~\citep{llm-bert}, \optbase{}~\citep{llm-opt}, \optlarge{}~\citep{llm-opt}, \llamayi{}~\citep{llm-llama-1}, \llamaer{}~\citep{llm-llama-2}, \llamasan{}~\citep{llm-llama-3}, \llamasy{}~\citep{llm-llama-3-1}, \chatgpt{}~\citep{llm-gpt35}, \qwensmall{}~\citep{llm-qwen}, \qwenbase{}~\citep{llm-qwen}, \qwen{}~\citep{llm-qwen}, \glm{}~\citep{llm-glm}, \mistral{}~\citep{llm-mistral}, \dsqwen{}~\citep{llm-deepseek}, \ewu{}~\citep{llm-e5}, \phier{}~\citep{llm-phi}, \recgpt{}~\citep{llm-recgpt}, \pwu{}~\citep{p5}, and \recformer{}~\citep{recformer}. We present a comprehensive comparison across multiple dimensions: varying model sizes within the same organization (e.g., Qwen-2 series), different versions from the same organization (e.g., Llama series), models of similar size released in the same year by different organizations (e.g., THU’s \glm{}, Meta’s \llamasan{}, and Alibaba’s \qwen{} in 2024), and models targeting different domains (e.g., the general foundation model Llama vs. the recommendation foundation model RecGPT). Specifically, the closed-source \chatgpt{} model from OpenAI supports only the prompt-based recommendation paradigm due to the unavailability of item and user embeddings. In contrast, models like \recformer{} and \ewu{}, designed with a dual-tower architecture, can only be evaluated with the embedding-based paradigm.

\textbf{Evaluation Protocols.}
Following common practice~\citep{bm-recbench}, we evaluate recommendation performance using widely adopted metrics, including ranking metrics such as \textsc{GAUC}, \textsc{nDCG}, and \textsc{MRR}, as well as matching metrics like \textsc{F1} and \textsc{Recall}. However, \textbf{due to space limitations}, we present only the \textsc{GAUC} (shortly \textsc{AUC}) metric mostly. The full evaluation results will available on our webpage.

\begin{figure*}
\centering
\resizebox{\linewidth}{!}{
\begin{tabular}{cc}
    \centering
    \begin{subfigure}[b]{0.48\textwidth} 
        \centering
        \includegraphics[width=\textwidth]{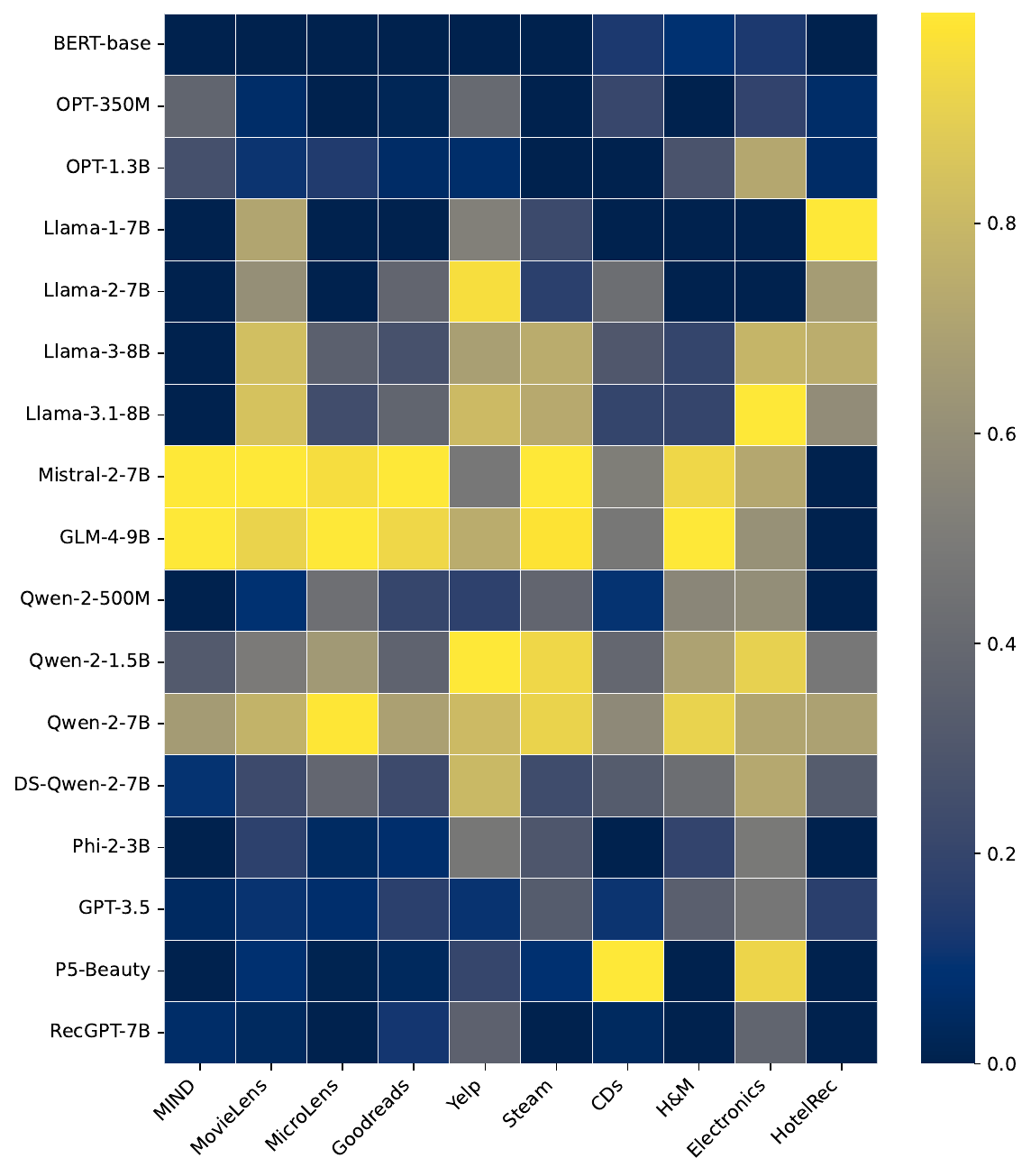}
        \caption{Prompt-based Evaluation}
        \label{fig:subfig1}
    \end{subfigure}
    \hspace{0.04\textwidth} 
    \begin{subfigure}[b]{0.48\textwidth}
        \centering
        \includegraphics[width=\textwidth]{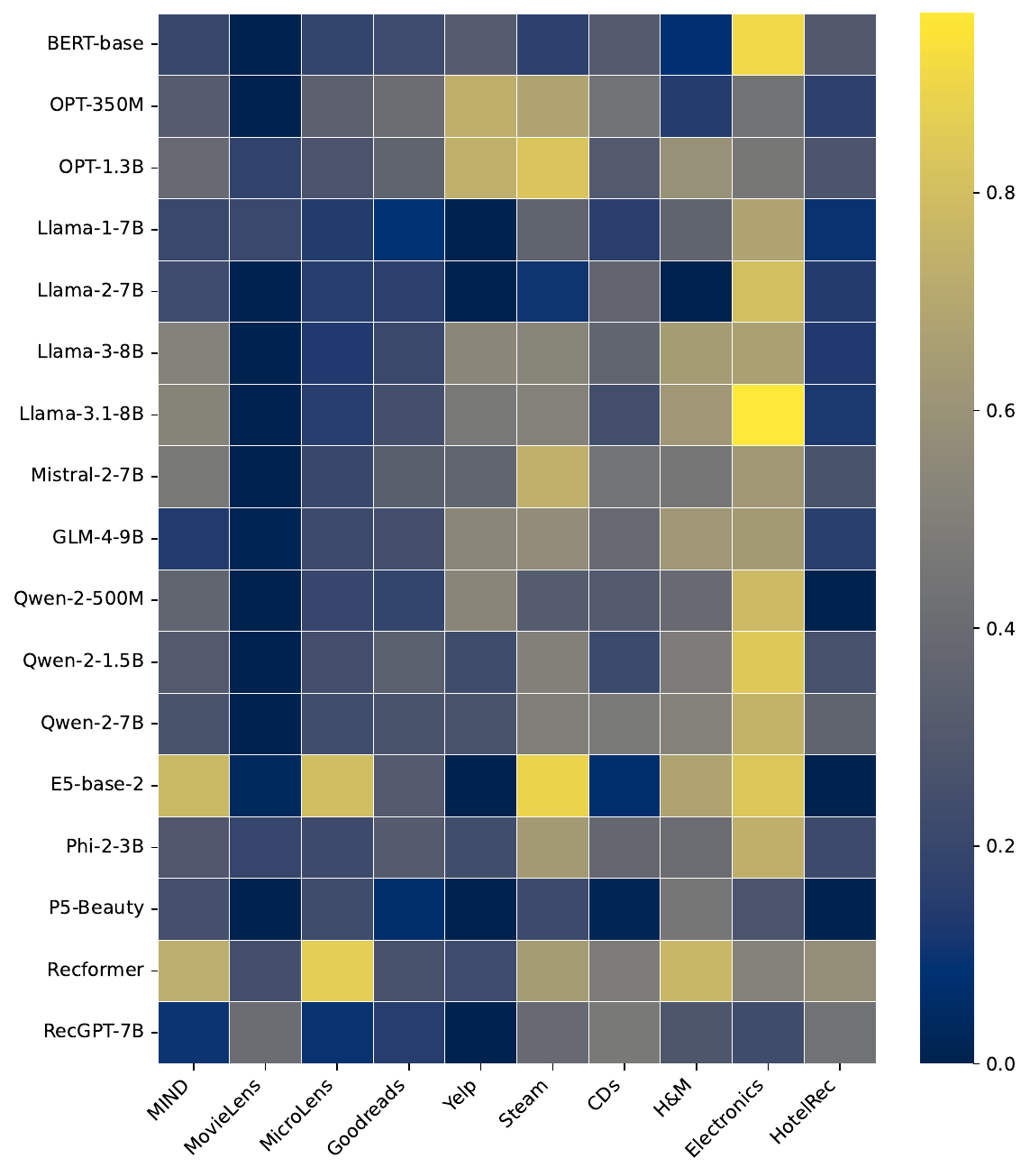}
        \caption{Embedding-based Evaluation}
        \label{fig:subfig2}
    \end{subfigure}
\end{tabular}
}
\caption{Zero-shot evaluation. The light-colored (yellow) areas indicate better \recabilities{}. Only AUC metric is reported due to the page limit.}
\label{fig:zero-shot}
\end{figure*}

Additionally, we also design the Reciprocal Rank Average (RRA) metric to evaluate the contribution for each finetune set (used in Table~\ref{tab:cross-dataset}). Specifically, we mark the top-K finetune set for each test set, and calculate the top-K RRA metric by:
\begin{gather}
\mathrm{RRA@K} = \frac{1}{T} \sum_{i=1}^{T} \left( \mathbb{I}(r_i <= K) \cdot \frac{1}{r_i} \right), \label{eq:rra} \\
\text{where indicator function } \mathbb{I}(b) = 1 \text{ if } b \text{ is true else } 0,
\end{gather}
where $T$ is the number of the test datasets, $r_i$ is the rank of the model on the $i$-th dataset, $K$ is the rank threshold (e.g., $K = 5$).

\textbf{Implementation Details.}
During data preprocessing, we standardized datasets of varying original sizes to comparable scales: the test set contains approximately 20,000 samples, while the fine-tuning set consists of around 100,000 samples. For each dataset, items were carefully curated to retain the most representative textual content features. User behavior sequences were truncated to a maximum length of 20; if a sequence exceeded this limit, only the most recent interactions were preserved.

We fine-tune models using LoRA~\citep{lora} (Low-Rank Adaptation), a parameter-efficient strategy with rank 32 and alpha 128. The learning rate is set to 1e-4 across all experiments, using the Adam optimizer. An effective batch size of 32 is maintained via gradient accumulation, and early stopping is applied with a patience of 2. Models are built and evaluated using the Huggingface Transformers library~\citep{transformers}. For \bertbase{}, \optlarge{}, and \llamasan{}, the maximum sequence lengths are 512, 1024, and 1024, respectively, with precision set to float32 for BERT and bfloat16 for \optlarge{} and \llamasan{}. To reduce fine-tuning overhead for embedding-based architectures, we freeze lower layers of \optlarge{} and \llamasan{}, applying LoRA only to the top two layers. When fine-tuning on multiple datasets, early stopping is based on the average validation AUC across datasets. We will release the code, data, checkpoints, and documentation at our GitHub repository.

All the experiments are conducted on a single Nvidia A100 GPU device. Except for the zero-shot setting, all results are averaged over five runs, with statistically significant differences observed ($p < 0.05$).

\section{Findings and Discussions}

In this section, we present a comprehensive analysis of experimental results evaluating the foundation model \recabilities{} in diverse fine-tuning regimes and different evaluation tasks.\footnote{Due to space limits, more experimental results are provided in the appendix and supplementary material.}

\begin{table}[]
\centering
\renewcommand{\arraystretch}{1.2}

\setlength\tabcolsep{3pt}

\caption{Performance comparison in single-domain fine-tuning scenario. We use cell background color to indicate different regimes, including \rB{}, \rC{}, and \rE{}. Only AUC metric is reported due to page limits.}
\label{tab:single-domain}

\resizebox{\linewidth}{!}{
\begin{tabular}{l|cc|cc|cc|cc|cc}
\toprule
\multirow{2}{*}{\makecell{\textbf{Foundation} \\ \textbf{Model}}}
 & \multicolumn{2}{c|}{\textbf{\hm{}}} 
 & \multicolumn{2}{c|}{\textbf{\mind{}}} 
 & \multicolumn{2}{c|}{\textbf{\microlens{}}} 
 & \multicolumn{2}{c|}{\textbf{\goodreads{}}} 
 & \multicolumn{2}{c}{\textbf{\cds{}}} \\
\cmidrule{2-11}
 & Prompt & Embedding & Prompt & Embedding & Prompt & Embedding & Prompt & Embedding & Prompt & Embedding \\
\midrule[1pt]
\textbf{Fine-tune set} & \multicolumn{10}{c}{\cellcolor{kbgB} N/A} \\
\midrule[1pt]
 \bertbase 
 & \cellcolor{kbgB} 0.5204 & \cellcolor{kbgB} 0.5167 
 & \cellcolor{kbgB} 0.4963 & \cellcolor{kbgB} 0.5263 
 & \cellcolor{kbgB} 0.4992 & \cellcolor{kbgB} 0.5305 
 & \cellcolor{kbgB} 0.4958 & \cellcolor{kbgB} 0.5160 
 & \cellcolor{kbgB} 0.5059 & \cellcolor{kbgB} 0.5139 
 \\
 
 \optlarge 
 & \cellcolor{kbgB} 0.5650 & \cellcolor{kbgB} 0.6370 
 & \cellcolor{kbgB} 0.5338 & \cellcolor{kbgB} 0.5510 
 & \cellcolor{kbgB} 0.5236 & \cellcolor{kbgB} 0.5447 
 & \cellcolor{kbgB} 0.5042 & \cellcolor{kbgB} 0.5258 
 & \cellcolor{kbgB} 0.4994 & \cellcolor{kbgB} 0.5137 
 \\
 
 \llamasan 
 & \cellcolor{kbgB} 0.5454 & \cellcolor{kbgB} 0.6487 
 & \cellcolor{kbgB} 0.4904 & \cellcolor{kbgB} 0.5666 
 & \cellcolor{kbgB} 0.5577 & \cellcolor{kbgB} 0.5218 
 & \cellcolor{kbgB} 0.5191 & \cellcolor{kbgB} 0.5150 
 & \cellcolor{kbgB} 0.5136 & \cellcolor{kbgB} 0.5162 
 \\
 
 \mistral 
 & \cellcolor{kbgB} 0.7166 & \cellcolor{kbgB} 0.6051 
 & \cellcolor{kbgB} 0.6300 & \cellcolor{kbgB} 0.5607 
 & \cellcolor{kbgB} 0.6579 & \cellcolor{kbgB} 0.5329 
 & \cellcolor{kbgB} 0.5718 & \cellcolor{kbgB} 0.5240 
 & \cellcolor{kbgB} 0.5230 & \cellcolor{kbgB} 0.5198 
 \\
 
 \qwen 
 & \cellcolor{kbgB} 0.7124 & \cellcolor{kbgB} 0.6201 
 & \cellcolor{kbgB} 0.5862 & \cellcolor{kbgB} 0.5347 
 & \cellcolor{kbgB} 0.6640 & \cellcolor{kbgB} 0.5391 
 & \cellcolor{kbgB} 0.5494 & \cellcolor{kbgB} 0.5190 
 & \cellcolor{kbgB} 0.5256 & \cellcolor{kbgB} 0.5212 
 \\
 
 \pwu 
 & \cellcolor{kbgB} 0.7124 & \cellcolor{kbgB} 0.6201 
 & \cellcolor{kbgB} 0.4911 & \cellcolor{kbgB} 0.5948 
 & \cellcolor{kbgB} 0.5017 & \cellcolor{kbgB} 0.6423 
 & \cellcolor{kbgB} 0.5027 & \cellcolor{kbgB} 0.5186 
 & \cellcolor{kbgB} 0.5447 & \cellcolor{kbgB} 0.5218 
 \\
 
\midrule[1pt]
\textbf{Fine-tune set} 
& \multicolumn{2}{c|}{\cellcolor{kbgC} \hm} 
& \multicolumn{2}{c|}{\cellcolor{kbgC} \mind} 
& \multicolumn{2}{c|}{\cellcolor{kbgC} \microlens} 
& \multicolumn{2}{c|}{\cellcolor{kbgC} \goodreads} 
& \multicolumn{2}{c}{\cellcolor{kbgC} \cds} 
\\
\midrule[1pt]

 \bertbase 
 & \cellcolor{kbgC} \textbf{0.8701} & \cellcolor{kbgC} 0.8359 
 & \cellcolor{kbgC} 0.7118 & \cellcolor{kbgC} 0.6837 
 & \cellcolor{kbgC} 0.8148 & \cellcolor{kbgC} \textbf{0.7671} 
 & \cellcolor{kbgC} 0.5208 & \cellcolor{kbgC} 0.5738 
 & \cellcolor{kbgC} 0.6185 & \cellcolor{kbgC} \textbf{0.5909} 
 \\
 
 \optlarge 
 & \cellcolor{kbgC} 0.5155 & \cellcolor{kbgC} 0.8163 
 & \cellcolor{kbgC} 0.7344 & \cellcolor{kbgC} 0.6788 
 & \cellcolor{kbgC} 0.7735 & \cellcolor{kbgC} 0.7545 
 & \cellcolor{kbgC} 0.4985 & \cellcolor{kbgC} 0.5754 
 & \cellcolor{kbgC} 0.5808 & \cellcolor{kbgC} 0.5713 
 \\
 
 \llamasan 
 & \cellcolor{kbgC} 0.8606 & \cellcolor{kbgC} 0.8150 
 & \cellcolor{kbgC} 0.7120 & \cellcolor{kbgC} 0.6745 
 & \cellcolor{kbgC} 0.8295 & \cellcolor{kbgC} 0.7430 
 & \cellcolor{kbgC} \textbf{0.6799} & \cellcolor{kbgC} \textbf{0.5944} 
 & \cellcolor{kbgC} 0.6267 & \cellcolor{kbgC} 0.5835 
 \\
 
\midrule[1pt]

\textbf{Fine-tune set} 

& \multicolumn{2}{c|}{\cellcolor{kbgE} \hm{} + \pog{}} 
& \multicolumn{2}{c|}{\cellcolor{kbgE} \mind{} + \pens{}} 
& \multicolumn{2}{c|}{\cellcolor{kbgE} \microlens{} + \netflix{}} 
& \multicolumn{2}{c|}{\cellcolor{kbgE} \goodreads{} + \books{}} 
& \multicolumn{2}{c}{\cellcolor{kbgE} \cds{} + \lastfm{}} 

\\
\midrule[1pt]

 \bertbase 
 & \cellcolor{kbgE} 0.8009 & \cellcolor{kbgE} \textbf{0.8373} 
 & \cellcolor{kbgE} 0.6834 & \cellcolor{kbgE} 0.6823 
 & \cellcolor{kbgE} 0.5480 & \cellcolor{kbgE} 0.7538 
 & \cellcolor{kbgE} 0.5169 & \cellcolor{kbgE} 0.5537 
 & \cellcolor{kbgE} 0.5790 & \cellcolor{kbgE} 0.5181 
 \\
 
 \optlarge 
 & \cellcolor{kbgE} 0.5208 & \cellcolor{kbgE} 0.7941 
 & \cellcolor{kbgE} \textbf{0.7350} & \cellcolor{kbgE} \textbf{0.6885} 
 & \cellcolor{kbgE} 0.5800 & \cellcolor{kbgE} 0.7443
 & \cellcolor{kbgE} 0.5966 & \cellcolor{kbgE} 0.5500 
 & \cellcolor{kbgE} 0.4926 & \cellcolor{kbgE} 0.5497 
 \\
 
 \llamasan 
 & \cellcolor{kbgE} 0.5378 & \cellcolor{kbgE} 0.8095 
 & \cellcolor{kbgE} \textbf{0.7350} & \cellcolor{kbgE} 0.6765 
 & \cellcolor{kbgE} \textbf{0.8368} & \cellcolor{kbgE} 0.7557
 & \cellcolor{kbgE} 0.6325 & \cellcolor{kbgE} 0.5638
 & \cellcolor{kbgE} \textbf{0.6353} & \cellcolor{kbgE} 0.5484
 \\
 
\bottomrule
\end{tabular}
}
\end{table}

\subsection{\rB{} Zero-shot Multi-domain: Prompt-based vs. Embedding-based}


Here, we investigate the zero-shot \recabilities{} of various foundation models.
For each dataset, we identify the maximum and minimum AUC values across all evaluated models in both paradigms (with the minimum constrained to 0.5) and normalize the results accordingly, as shown in Figure 2.
Based on these findings, we make the following observations:

First, for almost all datasets, the prompt-based evaluation paradigm outperforms the embedding-based one, as it aligns more closely with the pre-training objectives of foundation models.

Second, under the prompt-based paradigm, three LLMs (\mistral{}, \glm{}, \qwen{}) exhibit superior performance, possibly due to the inclusion of the collaborative signals during pre-training.
In contrast, \pwu{} performs well on two Amazon datasets (CDs and Electronics) but less favorably on others, likely because the used checkpoint was trained on the Amazon Beauty dataset, thereby modeling Amazon user interests.

Thirdly, in the embedding-based paradigm, performance differences among models are less pronounced.
Notably, \textbf{smaller models} (such as \bertbase{} and \optlarge{}) \textbf{perform better under this setting than in the prompt-based paradigm}, whereas the embeddings of larger models appear less sensitive to similarity metrics, in line with the findings of~\citep{embeddingrevisit}.
Additionally, the matching-based language model \ewu{} and recommendation model \recformer{} also demonstrate strong performance, benefiting from the consistency between the evaluation and training paradigms.

\subsection{Single-domain Fine-tuning: \rC{} vs. \rE{}}


Here, we study the single-domain fine-tuning recommendation scenario. We mainly select three foundation models, i.e., \bertbase, \optlarge, and \llamasan, of three distinct model size for the evaluation. From results displayed in Table~\ref{tab:single-domain}, we can make the following observations:

First, compared to zero-shot baselines (\rB{}), \textbf{domain-specific fine-tuning strategies} (\rC{} and \rE{}) consistently \textbf{achieve superior performance} on both prompt-based and embedding-based paradigm. This is primarily because large models have acquired domain-specific collaborative knowledge through fine-tuning.

Second, \textbf{fine-tuning with a single-domain single-dataset} setting (\rC{}) \textbf{yields more stable performance} than the cross-dataset variant (\rE{}), even within the same domain. This is likely due to optimization conflicts between datasets, as observed on Goodreads and H\&M, where \rE{} underperforms compared to \rC{}.

Third, \textbf{large-scale foundation models} (e.g., \llamasan) \textbf{achieve the best performance} under the \rE{}, as their pretraining enables a broad understanding of general textual knowledge across domains, allowing them to effectively extract and generalize useful patterns from auxiliary datasets to the target dataset. In contrast, smaller models such as \bertbase are less suited for \rE{}, as they struggle to abstract transferable patterns even from datasets within the same domain, leading to limited performance gains.

\begin{table}[]

\centering
\renewcommand{\arraystretch}{1.2}

\setlength\tabcolsep{1pt}

\caption{Performance comparison in cross-dataset fine-tuning scenario. We use cell background color to indicate different regimes, including \rB{}, \rC{}, \rD{}, and \rF{}. We mark the top-5 rank finetune set for each test set and bold the best. We use \textcolor{red}{\textbf{red}} color to indicate the result \textcolor{red}{\textbf{inferior}} to the zero-shot one or \textcolor{red}{\textbf{small}} than 0.5. Only AUC metric is reported due to page limits.}
\label{tab:cross-dataset}

\resizebox{\linewidth}{!}{
\begin{tabular}{l|llllllllll|l}
\toprule
 & \textbf{\hm{}} 
 & \textbf{\mind{}}
 & \textbf{\microlens{}}
 & \textbf{\goodreads{}} 
 & \textbf{\cds{}} 
 & \textbf{\movielens{}}
 & \textbf{\yelp{}} 
 & \textbf{\steam{}} 
 & \textbf{\electronics{}} 
 & \textbf{\hotelrec{}} 
 & \textbf{RRA@5} \\
\midrule[1pt]
\multicolumn{12}{c}{\cellcolor{bglight} \textbf{Foundation Model:} \bertbase} \\
\midrule[1pt]
N/A 
& \cellcolor{kbgB} 0.5204 & \cellcolor{kbgB} 0.4963 & \cellcolor{kbgB} 0.4992 
& \cellcolor{kbgB} 0.4958 & \cellcolor{kbgB} 0.5059 
& \cellcolor{kbgB} 0.4934 
& \cellcolor{kbgB} 0.4914 & \cellcolor{kbgB} 0.5002 
& \cellcolor{kbgB} 0.5037 & \cellcolor{kbgB} 0.4955 
& - \\
\midrule
\hm{}
& \cellcolor{kbgC} \textbf{0.8701} \small{(1)} 
& \cellcolor{kbgF} 0.5496 \small{(4)} 
& \cellcolor{kbgF} 0.5692 \small{(3)}
& \cellcolor{kbgF} \textbf{0.5282} \small{(1)} 
& \cellcolor{kbgF} 0.5103 \small{(3)}
& \cellcolor{kbgF} 0.5127 \small{(4)}
& \cellcolor{kbgF} \textcolor{red}{0.4961} 
& \cellcolor{kbgF} 0.7291 \small{(3)}
& \cellcolor{kbgF} 0.5304 \small{(3)} 
& \cellcolor{kbgF} \textcolor{red}{0.4869} 
& \textbf{0.3833} \small{(1)} \\
\mind{} 
& \cellcolor{kbgF} 0.6750 \small{(3)} 
& \cellcolor{kbgC} \textbf{0.7118} \small{(1)} 
& \cellcolor{kbgF} 0.5877 \small{(2)}
& \cellcolor{kbgF} 0.5255 \small{(4)} 
& \cellcolor{kbgF} 0.5128 \small{(2)}
& \cellcolor{kbgF} \textcolor{red}{0.4932}
& \cellcolor{kbgF} 0.5024 \small{(5)} 
& \cellcolor{kbgF} 0.7184 \small{(4)}
& \cellcolor{kbgF} 0.5306 \small{(2)} 
& \cellcolor{kbgF} \textcolor{red}{0.4847} 
& 0.3533 \small{(3)} \\
\microlens{} 
& \cellcolor{kbgF} 0.6661 \small{(4)} 
& \cellcolor{kbgF} 0.5841 \small{(3)} 
& \cellcolor{kbgC} \textbf{0.8148} \small{(1)}
& \cellcolor{kbgF} 0.5097 
& \cellcolor{kbgF} 0.5093 \small{(5)}
& \cellcolor{kbgF} 0.5150 \small{(3)}
& \cellcolor{kbgF} \textcolor{red}{0.4864} 
& \cellcolor{kbgF} \textbf{0.7393} \small{(1)}
& \cellcolor{kbgF} \textcolor{red}{0.5004}
& \cellcolor{kbgF} \textcolor{red}{0.4807}
& 0.3117 \small{(4)} \\
\goodreads{} 
& \cellcolor{kbgF} 0.6218
& \cellcolor{kbgF} 0.5081 
& \cellcolor{kbgF} 0.5239 
& \cellcolor{kbgC} 0.5208 \small{(5)} 
& \cellcolor{kbgF} \textcolor{red}{0.4992} 
& \cellcolor{kbgF} \textcolor{red}{0.4957}
& \cellcolor{kbgF} 0.5105 \small{(4)} 
& \cellcolor{kbgF} 0.6220 
& \cellcolor{kbgF} 0.5168 
& \cellcolor{kbgF} \textcolor{red}{0.4952} \small{(4)}
& 0.0700 \small{(8)} \\
\cds{} 
& \cellcolor{kbgF} 0.6464 \small{(5)} 
& \cellcolor{kbgF} 0.5053 
& \cellcolor{kbgF} 0.5152 
& \cellcolor{kbgF} 0.5139 
& \cellcolor{kbgC} \textbf{0.6185} \small{(1)}
& \cellcolor{kbgF} 0.5503 \small{(2)}
& \cellcolor{kbgF} \textbf{0.5356} \small{(1)} 
& \cellcolor{kbgF} \textcolor{red}{0.4794}
& \cellcolor{kbgF} 0.5076 
& \cellcolor{kbgF} \textbf{0.5216} \small{(1)}
& 0.3700 \small{(2)} \\
\pog{}
& \cellcolor{kbgD} 0.6222 
& \cellcolor{kbgF} 0.5153 
& \cellcolor{kbgF} 0.5470
& \cellcolor{kbgF} 0.5138 
& \cellcolor{kbgF} \textcolor{red}{0.4989}
& \cellcolor{kbgF} \textcolor{red}{0.4953}
& \cellcolor{kbgF} \textcolor{red}{0.4913} 
& \cellcolor{kbgF} 0.6171
& \cellcolor{kbgF} 0.5291 \small{(4)} 
& \cellcolor{kbgF} \textcolor{red}{0.4914} \small{(5)}
& 0.0450 (10) \\
\pens{}
& \cellcolor{kbgF} 0.6872 \small{(2)} 
& \cellcolor{kbgD} 0.6203 \small{(2)} 
& \cellcolor{kbgF} 0.5554 \small{(5)}
& \cellcolor{kbgF} 0.5165 
& \cellcolor{kbgF} 0.5051
& \cellcolor{kbgF} 0.5069
& \cellcolor{kbgF} \textcolor{red}{0.4987} 
& \cellcolor{kbgF} 0.7311 \small{(2)}
& \cellcolor{kbgF} 0.5218 
& \cellcolor{kbgF} \textcolor{red}{0.4900}
& 0.1700 \small{(7)} \\
\netflix{}
& \cellcolor{kbgF} 0.6191 
& \cellcolor{kbgF} 0.5396 \small{(5)} 
& \cellcolor{kbgD} 0.5328
& \cellcolor{kbgF} 0.5080 
& \cellcolor{kbgF} 0.5097 \small{(4)}
& \cellcolor{kbgF} \textbf{0.5656} \small{(1)}
& \cellcolor{kbgF} 0.5117 \small{(3)} 
& \cellcolor{kbgF} 0.6954 \small{(5)}
& \cellcolor{kbgF} 0.5255 \small{(5)} 
& \cellcolor{kbgF} 0.5077 \small{(2)}
& 0.2683 \small{(5)} \\
\books{}
& \cellcolor{kbgF} 0.6108 
& \cellcolor{kbgF} 0.5108 
& \cellcolor{kbgF} 0.5295
& \cellcolor{kbgD} 0.5264 \small{(2)} 
& \cellcolor{kbgF} 0.5089
& \cellcolor{kbgF} 0.5119 \small{(5)}
& \cellcolor{kbgF} 0.5155 \small{(2)} 
& \cellcolor{kbgF} 0.6191
& \cellcolor{kbgF} \textbf{0.5313} \small{(1)} 
& \cellcolor{kbgF} \textcolor{red}{0.4957} \small{(3)}
& 0.2533 \small{(6)} \\
\lastfm{}
& \cellcolor{kbgF} 0.6279 
& \cellcolor{kbgF} 0.5127 
& \cellcolor{kbgF} 0.5645 \small{(4)}
& \cellcolor{kbgF} 0.5263 \small{(3)} 
& \cellcolor{kbgD} \textcolor{red}{0.5023}
& \cellcolor{kbgF} \textcolor{red}{0.4855}
& \cellcolor{kbgF} \textcolor{red}{0.4773} 
& \cellcolor{kbgF} 0.6699
& \cellcolor{kbgF} 0.5231 
& \cellcolor{kbgF} \textcolor{red}{0.4769}
& 0.0583 \small{(9)} \\
\midrule[1pt]
\multicolumn{12}{c}{\cellcolor{bglight} \textbf{Foundation Model:} \llamasan} \\
\midrule[1pt]
N/A 
& \cellcolor{kbgB} 0.5267 & \cellcolor{kbgB} 0.4904 & \cellcolor{kbgB} 0.6412 
& \cellcolor{kbgB} 0.5577 & \cellcolor{kbgB} 0.5191 
& \cellcolor{kbgB} 0.7690 
& \cellcolor{kbgB} 0.5136 & \cellcolor{kbgB} 0.5454 
& \cellcolor{kbgB} 0.5223 & \cellcolor{kbgB} 0.5342 
& - \\
\midrule
\hm{}
& \cellcolor{kbgC} \textbf{0.8606} \small{(1)} 
& \cellcolor{kbgF} 0.5693 \small{(5)} 
& \cellcolor{kbgF} 0.6758 \small{(2)}
& \cellcolor{kbgF} 0.6116 \small{(2)} 
& \cellcolor{kbgF} 0.5268 \small{(5)}
& \cellcolor{kbgF} \textcolor{red}{0.6818} \small{(4)}
& \cellcolor{kbgF} \textcolor{red}{0.4911} 
& \cellcolor{kbgF} 0.9193 \small{(2)}
& \cellcolor{kbgF} 0.5469 \small{(4)} 
& \cellcolor{kbgF} \textcolor{red}{0.5116}
& 0.3400 \small{(2)} \\
\mind{}
& \cellcolor{kbgF} 0.6599 
& \cellcolor{kbgC} \textbf{0.7120} \small{(1)} 
& \cellcolor{kbgF} \textcolor{red}{0.6104} \small{(5)}
& \cellcolor{kbgF} \textcolor{red}{0.5279} 
& \cellcolor{kbgF} \textcolor{red}{0.5176}

& \cellcolor{kbgF} \textcolor{red}{0.5235}
& \cellcolor{kbgF} \textcolor{red}{0.4808} 
& \cellcolor{kbgF} 0.8033
& \cellcolor{kbgF} 0.5284 
& \cellcolor{kbgF} \textcolor{red}{0.5008}
& 0.1200 \small{(9)} \\
\microlens{}
& \cellcolor{kbgF} 0.7504 \small{(3)} 
& \cellcolor{kbgF} 0.6331 \small{(3)} 
& \cellcolor{kbgC} \textbf{0.8295} \small{(1)}
& \cellcolor{kbgF} 0.5829 
& \cellcolor{kbgF} 0.5239

& \cellcolor{kbgF} \textcolor{red}{0.6703}
& \cellcolor{kbgF} 0.5165 
& \cellcolor{kbgF} 0.8953 \small{(4)}
& \cellcolor{kbgF} 0.5250 
& \cellcolor{kbgF} \textcolor{red}{0.4900}
& 0.1917 \small{(7)} \\

\goodreads{}
& \cellcolor{kbgF} 0.6592 
& \cellcolor{kbgF} 0.5249 
& \cellcolor{kbgF} \textcolor{red}{0.5731}
& \cellcolor{kbgC} \textbf{0.6799} \small{(1)} 
& \cellcolor{kbgF} 0.5334 \small{(4)}

& \cellcolor{kbgF} \textcolor{red}{0.6550}
& \cellcolor{kbgF} 0.5301 \small{(3)} 
& \cellcolor{kbgF} 0.8795 \small{(5)}
& \cellcolor{kbgF} 0.6051 \small{(3)} 
& \cellcolor{kbgF} \textcolor{red}{0.5320} \small{(4)}
& 0.2367 \small{(5)} \\
\cds{}
& \cellcolor{kbgF} 0.6262 
& \cellcolor{kbgF} 0.5019 
& \cellcolor{kbgF} \textcolor{red}{0.5385}
& \cellcolor{kbgF} 0.5840 \small{(5)} 
& \cellcolor{kbgC} \textbf{0.6267} \small{(1)}

& \cellcolor{kbgF} \textcolor{red}{0.7201} \small{(2)}
& \cellcolor{kbgF} \textbf{0.5939} \small{(1)} 
& \cellcolor{kbgF} 0.6427
& \cellcolor{kbgF} \textbf{0.6410} \small{(1)} 
& \cellcolor{kbgF} 0.6057 \small{(3)}
& \textbf{0.4033} \small{(1)} \\

\pog{}
& \cellcolor{kbgD} 0.5922 
& \cellcolor{kbgF} \textcolor{red}{0.4912} 
& \cellcolor{kbgF} \textcolor{red}{0.5175}
& \cellcolor{kbgF} \textcolor{red}{0.5068} 
& \cellcolor{kbgF} \textcolor{red}{0.5011}

& \cellcolor{kbgF} \textcolor{red}{0.5354}
& \cellcolor{kbgF} 0.5147 \small{(4)} 
& \cellcolor{kbgF} 0.5684
& \cellcolor{kbgF} \textcolor{red}{0.4896} 
& \cellcolor{kbgF} \textcolor{red}{0.5142} \small{(5)}
& 0.0450 (10) \\

\pens{}
& \cellcolor{kbgF} 0.7517 \small{(2)} 
& \cellcolor{kbgD} 0.6823 \small{(2)} 
& \cellcolor{kbgF} 0.6675 \small{(3)}
& \cellcolor{kbgF} 0.5670 
& \cellcolor{kbgF} 0.5191 

& \cellcolor{kbgF} \textcolor{red}{0.6311}
& \cellcolor{kbgF} \textcolor{red}{0.5134} \small{(5)} 
& \cellcolor{kbgF} 0.8979 \small{(3)}
& \cellcolor{kbgF} 0.5353 
& \cellcolor{kbgF} \textcolor{red}{0.4860}
& 0.1867 \small{(8)} \\

\netflix{}
& \cellcolor{kbgF} 0.6655 \small{(5)} 
& \cellcolor{kbgF} \textcolor{red}{0.4844} 
& \cellcolor{kbgD} \textcolor{red}{0.5634}
& \cellcolor{kbgF} 0.5694 
& \cellcolor{kbgF} 0.5355 \small{(3)}

& \cellcolor{kbgF} \textbf{\textcolor{red}{0.7422}} \small{(1)}
& \cellcolor{kbgF} \textcolor{red}{0.4958} 
& \cellcolor{kbgF} 0.8547 
& \cellcolor{kbgF} 0.5947 \small{(3)} 
& \cellcolor{kbgF} 0.6104 \small{(2)}
& 0.2367 \small{(5)} \\

\books{}
& \cellcolor{kbgF} 0.5750 
& \cellcolor{kbgF} 0.5040 
& \cellcolor{kbgF} \textcolor{red}{0.5322}
& \cellcolor{kbgD} 0.5876 \small{(3)} 
& \cellcolor{kbgF} 0.5600 \small{(2)}

& \cellcolor{kbgF} \textcolor{red}{0.6935} \small{(3)}
& \cellcolor{kbgF} 0.5727 \small{(2)} 
& \cellcolor{kbgF} 0.6373
& \cellcolor{kbgF} 0.6267 \small{(2)} 
& \cellcolor{kbgF} \textbf{0.6149} \small{(1)}
& 0.2833 \small{(3)} \\

\lastfm{}
& \cellcolor{kbgF} 0.7444 \small{(4)} 
& \cellcolor{kbgF} 0.5807 \small{(4)} 
& \cellcolor{kbgF} 0.6567 \small{(4)}
& \cellcolor{kbgF} 0.5871 \small{(4)} 
& \cellcolor{kbgD} \textcolor{red}{0.5045}

& \cellcolor{kbgF} \textcolor{red}{0.6736} \small{(5)}
& \cellcolor{kbgF} \textcolor{red}{0.4818} 
& \cellcolor{kbgF} \textbf{0.9363} \small{(1)}
& \cellcolor{kbgF} 0.5414 \small{(5)} 
& \cellcolor{kbgF} \textcolor{red}{0.5125}
& 0.2400 \small{(4)} \\
\bottomrule
\end{tabular}
}  
\end{table}

\subsection{Cross-dataset Fine-tuning: \rD{} vs. \rF{}}

Here, we study effect of the cross-dataset fine-tuning, including single-domain and cross-domain scenario. The experiments are conducted across two foundation models: \bertbase{} and \llamasan{}. The foundation model will be firstly fine-tuned with one single dataset and then evaluated over 10 test datasets. We design a RRA metric (Equation~\ref{eq:rra}) to evaluate the usefulness of each finetune set.

Based Table~\ref{tab:cross-dataset}, we can make the following observations:

First, \textbf{cross-dataset fine-tuning} generally improves recommendation performance, but it \textbf{may also introduce negative effects} on the target dataset in some cases (as indicated by the red-highlighted results in the table). Notably, the \yelp{} and \hotelrec{} datasets exhibit a higher likelihood of such degradation, possibly due to domain gaps and mismatches between the test sets and finetune sets.
Moreover, for the \llamasan{} model, \microlens{} and \movielens{} also demonstrate performance degradation under cross-dataset finetuning. Interestingly, these two datasets are where \llamasan{} achieves the highest zero-shot performance among the 10 test sets. This suggests that \llamasan{} likely encountered collaborative signals related to these domains during pretraining, allowing it to effectively capture user interests for video-based recommendations even without additional tuning.

Second, \textbf{single-domain cross-dataset finetuning is not always more effective than cross-domain finetuning}. While it intuitively makes sense that user interests are easier to model within the same domain--supported by results in news (\mind{}--\pens{}) and books (\goodreads{}--\books{})--this trend does not hold for movies, music, and fashion: their \rD{} results did not even rank in the top five. A possible reason is that \mind{} and \pens{} both originate from Microsoft, and Amazon is the source of \books{} as well as the parent company of Goodreads.com, suggesting these dataset pairs may share more similar distributions.

Third, \textbf{dataset quality varies, but its effectiveness also depends heavily on the capacity of the pretrained model}. For instance, finetuning on \goodreads{} with \bertbase{} (\rC{}) ranks only fifth, while using \llamasan{} lifts it to first. This may be because Goodreads relies on book titles as content features, which are poorly represented in smaller models’ pretraining corpora. In contrast, \llamasan{} better understands textual content, leading to more robust item representations and improved user modeling. According to the Top-5 RRA results, \cds{} and \hm{} offer the strongest transferability, while \pog{} performs the weakest. Additionally, \goodreads{} and \lastfm{} show large performance gains when switching from \bertbase{} to \llamasan{}, suggesting complex content features paired with highly transferable user interests. On the other hand, \mind{} and \microlens{} show ranking drops, indicating their simpler content may already be sufficiently modeled by smaller models, but their user behavior patterns are less suitable for cross-dataset transfer.
Finally, although \books{} and \lastfm{} do not have corresponding test sets, their finetuned models still rank in the top five under \llamasan{}, suggesting strong generalization capability across domains.

\subsection{Multi-domain Fine-tuning: \rG{} vs. \rH{}}

We investigate the impact of multi-domain fine-tuning, focusing on two key settings: multi-domain cross-dataset (\rG{}) and multi-domain multi-dataset (\rH{}). In the \rG{} setting, foundation models are fine-tuned using datasets from domains different from the test sets, specifically: \pog{}, \pens{}, \netflix{}, \books{}, and \lastfm{}. In contrast, the \rH{} setting involves fine-tuning on datasets that share domains with the test sets but do not include overlapping data, namely: \hm{}, \mind{}, \microlens{}, \goodreads{}, and \cds{}. From the results illustrated in Table~\ref{tab:multi-domain}, we can make the following observations:

First, \rH{} achieves the best performance on \hm{}, \mind{}, \microlens{}, \goodreads{}, and \cds{} across all three foundation models, as it is fine-tuned directly on these datasets and thus captures domain-specific knowledge effectively. Second, although \rG{} uses different datasets for fine-tuning, it consistently outperforms the zero-shot setting (\rB{}), highlighting \textbf{the generalization benefits of multi-domain training with diverse user behavior patterns}. Third, while \rH{} serves as an upper bound, the performance gap between \rH{} and \rG{} narrows with larger models--for instance, the improvement on HM drops from 30.0\% (\bertbase{}) to 16.8\% (\llamasan{}), suggesting that large models fine-tuned on cross-domain data can better handle zero-resource scenarios. Finally, due to its reliance on specific data distributions, \rH{} underperforms \rG{} on five other datasets (\movielens{}, \yelp{}, \steam{}, \electronics{}, \hotelrec{}), underscoring the fairness and robustness of the multi-domain cross-dataset setting.

\begin{table}[]

\centering
\renewcommand{\arraystretch}{1.2} 

\setlength\tabcolsep{3pt}

\caption{Performance comparison in multi-domain fine-tuning scenario. We use cell background color to indicate different settings, including \rB{}, \rG{}, and \rH{}. Only AUC metric is reported due to page limits.}\label{tab:multi-domain}

\resizebox{\linewidth}{!}{

\begin{tabular}{l|cccccccccc} 
\toprule
Model & \textbf{\hm} & \textbf{\mind} & \textbf{\microlens} & \textbf{\goodreads} &\textbf{ \cds }& \textbf{\movielens} & \textbf{\yelp} & \textbf{\steam} & \textbf{\electronics }& \textbf{\hotelrec} \\
\midrule
\multirow{3}{*}{\bertbase} & \cellcolor{kbgB} 0.5204 & \cellcolor{kbgB} 0.4963 & \cellcolor{kbgB} 0.4992 & \cellcolor{kbgB} 0.4958 & \cellcolor{kbgB} 0.5059 & \cellcolor{kbgB} 0.4934 & \cellcolor{kbgB} 0.4914 & \cellcolor{kbgB} 0.5002 & \cellcolor{kbgB} 0.5037 & \cellcolor{kbgB} \textbf{0.4955} \\
 & \cellcolor{kbgG} 0.6607 & \cellcolor{kbgG} 0.6004 & \cellcolor{kbgG} 0.5711 & \cellcolor{kbgG} 0.5213 & \cellcolor{kbgG} 0.5092 & \cellcolor{kbgG} \textbf{0.5675} & \cellcolor{kbgG} \textbf{0.5218} & \cellcolor{kbgG} 0.6189 & \cellcolor{kbgG} \textbf{0.5091} & \cellcolor{kbgG} 0.4795 \\
 & \cellcolor{kbgH} \textbf{0.8569} & \cellcolor{kbgH} \textbf{0.7004} & \cellcolor{kbgH} \textbf{0.8019} & \cellcolor{kbgH} \textbf{0.5648} & \cellcolor{kbgH} \textbf{0.5653} & \cellcolor{kbgH} 0.5106 & \cellcolor{kbgH} 0.5093 & \cellcolor{kbgH} \textbf{0.6817} & \cellcolor{kbgH} 0.5039 & \cellcolor{kbgH} 0.4869 \\
\midrule
\multirow{3}{*}{\optlarge} & \cellcolor{kbgB} 0.5650 & \cellcolor{kbgB} 0.5338 & \cellcolor{kbgB} 0.5236 & \cellcolor{kbgB} 0.5042 & \cellcolor{kbgB} 0.4994 & \cellcolor{kbgB} 0.5174 & \cellcolor{kbgB} \textbf{0.5026} & \cellcolor{kbgB} 0.3825 & \cellcolor{kbgB} 0.5205 & \cellcolor{kbgB} 0.5026 \\
 & \cellcolor{kbgG} 0.7002 & \cellcolor{kbgG} 0.5996 & \cellcolor{kbgG} 0.6165 & \cellcolor{kbgG} 0.5189 & \cellcolor{kbgG} 0.5181 & \cellcolor{kbgG} \textbf{0.6156} & \cellcolor{kbgG} 0.4853 & \cellcolor{kbgG} \textbf{0.7665} & \cellcolor{kbgG} \textbf{0.5446} & \cellcolor{kbgG} \textbf{0.5037} \\
 & \cellcolor{kbgH} \textbf{0.8658} & \cellcolor{kbgH} \textbf{0.7259} & \cellcolor{kbgH} \textbf{0.8132} & \cellcolor{kbgH} \textbf{0.5374} & \cellcolor{kbgH} \textbf{0.6220} & \cellcolor{kbgH} 0.5813 & \cellcolor{kbgH} 0.5014 & \cellcolor{kbgH} 0.6959 & \cellcolor{kbgH} 0.5248 & \cellcolor{kbgH} 0.4951 \\
\midrule
\multirow{3}{*}{\llamasan} & \cellcolor{kbgB} 0.7690 & \cellcolor{kbgB} 0.4904 & \cellcolor{kbgB} 0.6412 & \cellcolor{kbgB} 0.5577 & \cellcolor{kbgB} 0.5191 & \cellcolor{kbgB} 0.5267 & \cellcolor{kbgB} 0.5136 & \cellcolor{kbgB} 0.5454 & \cellcolor{kbgB} 0.5223 & \cellcolor{kbgB} 0.5342 \\
 & \cellcolor{kbgG} 0.7295 & \cellcolor{kbgG} 0.6732 & \cellcolor{kbgG} 0.6223 & \cellcolor{kbgG} 0.5864 & \cellcolor{kbgG} 0.5626 & \cellcolor{kbgG} \textbf{0.7203} & \cellcolor{kbgG} 0.5764 & \cellcolor{kbgG} \textbf{0.7828} & \cellcolor{kbgG} \textbf{0.6296} & \cellcolor{kbgG} \textbf{0.5806} \\
 & \cellcolor{kbgH} \textbf{0.8524} & \cellcolor{kbgH} \textbf{0.7206} & \cellcolor{kbgH} \textbf{0.8235} & \cellcolor{kbgH} \textbf{0.6660} & \cellcolor{kbgH} \textbf{0.6281} & \cellcolor{kbgH} 0.6683 & \cellcolor{kbgH} \textbf{0.5846} & \cellcolor{kbgH} 0.7042 & \cellcolor{kbgH} 0.6139 & \cellcolor{kbgH} 0.5318 \\
\bottomrule
\end{tabular}
}
\end{table}

\section{Conclusion}

We have introduced \recbenchmd{}, a novel and comprehensive benchmark designed to evaluate the recommendation capabilities of foundation models across a wide range of datasets and domains. Our thorough analysis of 19 foundation models across 15 datasets and 10 domains provides crucial insights into their performance in recommendation tasks. The findings demonstrate the substantial advantages of cross-dataset transfer learning and multi-domain training in improving the adaptability of foundation models. We expect that these insights, along with the valuable resources provided, will drive future advancements in the development of recommendation foundation models, offering a strong foundation for continued research and innovation in this field.


\bibliographystyle{plainnat}
\bibliography{RecBench-MD}

\appendix

\section{Limitations} \label{sec:limitation}

In this study, we assess the recommendation capabilities of foundation models on two of the most prevalent tasks: prompt-based approaches (similar to CTR models) and embedding-based approaches (akin to matching models), from a multi-dataset, multi-domain perspective. Nonetheless, our current evaluation does not encompass sequential recommendation, which represents a crucial area for future development and enhancement.


\section{Broader Impacts}


Our benchmark offers a comprehensive and scalable framework for evaluating foundation models in zero-resource, multi-dataset, multi-domain recommendation scenarios, thereby promoting more systematic and reproducible research. It establishes a solid foundation for ongoing research and innovation in this field. Furthermore, the benchmark facilitates cross-domain fine-tuning, extending its benefits to other areas such as natural language processing.

\newpage

\section{Technical Appendices}

\begin{table}[t]

\centering
\renewcommand{\arraystretch}{1.2} 

\setlength\tabcolsep{3pt}

\caption{\textbf{Performance comparison across various fine-tuning datasets and orders}. Each sub-table presents 25 AUC scores on a test set. For the entry at row $i$, column $j$ (e.g., $i=1$, $j=2$ in the \mind{} sub-table), the model is first fine-tuned on \pog{}, then on \pens{}, yielding an AUC of 0.6687 when tested on \mind{}. ``Overall'' indicates the average performance across the corresponding five test sets. The diagonal cells (highlighted in grey) represent results of single-dataset fine-tuning. We rank these five values and annotate the rank next to each dataset name in the column header (e.g., \pens{} (1) in the \mind{} sub-table indicates that 0.6823 is the highest result in single-dataset fine-tuning). For each pair of datasets, different fine-tuning orders generally yield significantly different results. The superior result for each pair is highlighted in green. The top five results among the 25 entries within each sub-table are annotated with their respective rankings.
}\label{tab:order}

\resizebox{\linewidth}{!}{
\begin{tabular}{l|lllll|lllll}
\toprule
\midrule
 &
  \multicolumn{5}{c|}{\hm} &
  \multicolumn{5}{c}{\mind} \\
\midrule[1pt]
 &
  \pog{} (4) &
  \pens{} (1) &
  \netflix{} (3) &
  \books{} (5) &
  \lastfm{} (2) &
  \pog{} (4) &
  \pens{} (1) &
  \netflix{} (5) &
  \books{} (3) &
  \lastfm{} (2) \\
\midrule
\pog &
  \cellcolor[HTML]{d5d5d5} 0.5922 &
  \cellcolor{bgGood} 0.7287 &
  \cellcolor[HTML]{ffffff} 0.6722 &
  \cellcolor[HTML]{ffffff} 0.5824 &
  \cellcolor{bgGood} 0.7496 (4) &
  \cellcolor[HTML]{d5d5d5} 0.4912 &
  \cellcolor{bgGood} 0.6687 (5) &
  \cellcolor[HTML]{ffffff} 0.5221 &
  \cellcolor[HTML]{ffffff} 0.4863 &
  \cellcolor{bgGood} 0.5901 \\
\pens &
  \cellcolor[HTML]{ffffff} 0.6827 &
  \cellcolor[HTML]{d5d5d5} 0.7517 (2) &
  \cellcolor[HTML]{ffffff} 0.7338 &
  \cellcolor[HTML]{ffffff} 0.6036 &
  \cellcolor[HTML]{ffffff} 0.7355 &
  \cellcolor[HTML]{ffffff} 0.6636 &
  \cellcolor[HTML]{d5d5d5} 0.6823 (1) &
  \cellcolor[HTML]{ffffff} 0.6186 &
  \cellcolor[HTML]{ffffff} 0.5270 &
  \cellcolor[HTML]{ffffff} 0.6482 \\
\netflix &
  \cellcolor{bgGood} 0.7099 &
  \cellcolor{bgGood} 0.7439 &
  \cellcolor[HTML]{d5d5d5} 0.6655 &
  \cellcolor[HTML]{ffffff} 0.6442 &
  \cellcolor{bgGood} 0.7502 (3) &
  \cellcolor{bgGood} 0.5530 &
  \cellcolor{bgGood} 0.6757 (3) &
  \cellcolor[HTML]{d5d5d5} 0.4844 &
  \cellcolor[HTML]{ffffff} 0.4992 &
  \cellcolor{bgGood} 0.5866 \\
\books &
  \cellcolor{bgGood} 0.7356 &
  \cellcolor{bgGood} 0.7450 (5) &
  \cellcolor{bgGood} 0.6923 &
  \cellcolor[HTML]{d5d5d5} 0.5750 &
  \cellcolor{bgGood} 0.7444 &
  \cellcolor{bgGood} 0.5248 &
  \cellcolor{bgGood} 0.6759 (2) &
  \cellcolor{bgGood} 0.5223 &
  \cellcolor[HTML]{d5d5d5} 0.5040 &
  \cellcolor{bgGood} 0.5785 \\
\lastfm &
  \cellcolor[HTML]{ffffff} 0.7367 &
  \cellcolor{bgGood} 0.7592 (1) &
  \cellcolor[HTML]{ffffff} 0.7143 &
  \cellcolor[HTML]{ffffff} 0.6096 &
  \cellcolor[HTML]{d5d5d5} 0.7444 &
  \cellcolor[HTML]{ffffff} 0.5440 &
  \cellcolor{bgGood} 0.6738 (4) &
  \cellcolor[HTML]{ffffff} 0.5039 &
  \cellcolor[HTML]{ffffff} 0.5149 &
  \cellcolor[HTML]{d5d5d5} 0.5807 \\
\midrule
 &
  \multicolumn{5}{c|}{\microlens} &
  \multicolumn{5}{c}{\goodreads} \\
\midrule
 &
  \pog{} (5) &
  \pens{} (1) &
  \netflix{} (3) &
  \books{} (4) &
  \lastfm{} (2) &
  \pog{} (5) &
  \pens{} (4) &
  \netflix{} (3) &
  \books{} (1) &
  \lastfm{} (2) \\
\pog &
  \cellcolor[HTML]{d5d5d5} 0.5175 &
  \cellcolor{bgGood} 0.6443 &
  \cellcolor[HTML]{ffffff} 0.5753 &
  \cellcolor[HTML]{ffffff} 0.5304 &
  \cellcolor{bgGood} 0.6495 &
  \cellcolor[HTML]{d5d5d5} 0.5068 &
  \cellcolor{bgGood} 0.5659 &
  \cellcolor[HTML]{ffffff} 0.5631 &
  \cellcolor[HTML]{ffffff} 0.5700 &
  \cellcolor{bgGood} 0.5858 \\
\pens &
  \cellcolor[HTML]{ffffff} 0.6262 &
  \cellcolor[HTML]{d5d5d5} 0.6675 &
  \cellcolor[HTML]{ffffff} 0.6157 &
  \cellcolor[HTML]{ffffff} 0.5508 &
  \cellcolor[HTML]{ffffff} 0.6439 &
  \cellcolor[HTML]{ffffff} 0.5632 &
  \cellcolor[HTML]{d5d5d5} 0.5670 &
  \cellcolor{bgGood} 0.5810 &
  \cellcolor{bgGood} 0.6024 (1) &
  \cellcolor[HTML]{ffffff} 0.5375 \\
\netflix &
  \cellcolor{bgGood} 0.6222 &
  \cellcolor{bgGood} 0.6693 (2) &
  \cellcolor[HTML]{d5d5d5} 0.5634 &
  \cellcolor[HTML]{ffffff} 0.5650 &
  \cellcolor{bgGood} 0.6688 (3) &
  \cellcolor{bgGood} 0.5690 &
  \cellcolor[HTML]{ffffff} 0.5721 &
  \cellcolor[HTML]{d5d5d5} 0.5694 &
  \cellcolor{bgGood} 0.5973 (2) &
  \cellcolor[HTML]{ffffff} 0.5743 \\
\books &
  \cellcolor{bgGood} 0.6328 &
  \cellcolor{bgGood} 0.6662 (4) &
  \cellcolor{bgGood} 0.5784 &
  \cellcolor[HTML]{d5d5d5} 0.5322 &
  \cellcolor{bgGood} 0.6462 &
  \cellcolor{bgGood} 0.5822 &
  \cellcolor[HTML]{ffffff} 0.5947 (5) &
  \cellcolor[HTML]{ffffff} 0.5950 (4) &
  \cellcolor[HTML]{d5d5d5} 0.5876 &
  \cellcolor[HTML]{ffffff} 0.5864 \\
\lastfm &
  \cellcolor[HTML]{ffffff} 0.6363 &
  \cellcolor{bgGood} 0.6779 (1) &
  \cellcolor[HTML]{ffffff} 0.5875 &
  \cellcolor[HTML]{ffffff} 0.5338 &
  \cellcolor[HTML]{d5d5d5} 0.6567 (5) &
  \cellcolor[HTML]{ffffff} 0.5540 &
  \cellcolor{bgGood} 0.5728 &
  \cellcolor{bgGood} 0.5748 &
  \cellcolor{bgGood} 0.5953 (3) &
  \cellcolor[HTML]{d5d5d5} 0.5871 \\
\midrule 
 &
  \multicolumn{5}{c|}{\cds} &
  \multicolumn{5}{c}{Overall} \\
\midrule
 &
  \pog{} (5) &
  \pens{} (3) &
  \netflix{} (2) &
  \books{} (1) &
  \lastfm{} (4) &
  \pog{} (5) &
  \pens{} (1) &
  \netflix{} (3) &
  \books{} (4) &
  \lastfm{} (2) \\
\pog &
  \cellcolor[HTML]{d5d5d5} 0.5011 &
  \cellcolor{bgGood} 0.5225 &
  \cellcolor{bgGood} 0.5254 &
  \cellcolor[HTML]{ffffff} 0.5517 &
  \cellcolor[HTML]{ffffff} 0.4947 &
  \cellcolor[HTML]{d5d5d5} 0.5218 &
  \cellcolor{bgGood} 0.6260 (5) &
  \cellcolor[HTML]{ffffff} 0.5716 &
  \cellcolor[HTML]{ffffff} 0.5442 &
  \cellcolor{bgGood} 0.6139 \\
\pens &
  \cellcolor[HTML]{ffffff} 0.5140 &
  \cellcolor[HTML]{d5d5d5} 0.5191 &
  \cellcolor{bgGood} 0.5366 &
  \cellcolor{bgGood} 0.5551 (5) &
  \cellcolor[HTML]{ffffff} 0.5023 &
  \cellcolor[HTML]{ffffff} 0.6099 &
  \cellcolor[HTML]{d5d5d5} 0.6375 (4) &
  \cellcolor[HTML]{ffffff} 0.6171 &
  \cellcolor[HTML]{ffffff} 0.5678 &
  \cellcolor[HTML]{ffffff} 0.6135 \\
\netflix &
  \cellcolor[HTML]{ffffff} 0.5237 &
  \cellcolor[HTML]{ffffff} 0.5293 &
  \cellcolor[HTML]{d5d5d5} 0.5355 &
  \cellcolor{bgGood} 0.5602 (2) &
  \cellcolor[HTML]{ffffff} 0.5098 &
  \cellcolor{bgGood} 0.5956 &
  \cellcolor{bgGood} 0.6381 (3) &
  \cellcolor[HTML]{d5d5d5} 0.5636 &
  \cellcolor[HTML]{ffffff} 0.5732 &
  \cellcolor{bgGood} 0.6179 \\
\books &
  \cellcolor{bgGood} 0.5563 (4) &
  \cellcolor[HTML]{ffffff} 0.5340 &
  \cellcolor[HTML]{ffffff} 0.5387 &
  \cellcolor[HTML]{d5d5d5} 0.5600 (3) &
  \cellcolor[HTML]{ffffff} 0.5039 &
  \cellcolor{bgGood} 0.6063 &
  \cellcolor{bgGood} 0.6432 (1) &
  \cellcolor{bgGood} 0.5853 &
  \cellcolor[HTML]{d5d5d5} 0.5518 &
  \cellcolor{bgGood} 0.6119 \\
\lastfm &
  \cellcolor{bgGood} 0.5098 &
  \cellcolor{bgGood} 0.5263 &
  \cellcolor{bgGood} 0.5259 &
  \cellcolor{bgGood} 0.5654 (1) &
  \cellcolor[HTML]{d5d5d5} 0.5045 &
  \cellcolor[HTML]{ffffff} 0.5962 &
  \cellcolor{bgGood} 0.6420 (2) &
  \cellcolor[HTML]{ffffff} 0.5813 &
  \cellcolor[HTML]{ffffff} 0.5638 &
  \cellcolor[HTML]{d5d5d5} 0.6147 \\

\midrule
\midrule
 & \multicolumn{5}{c|}{\movielens{}} & \multicolumn{5}{c}{\yelp{}} \\
\midrule[1pt]
 & \pog{} (5) & \pens{} (4) & \netflix{} (1) & \books{} (2) & \lastfm{} (3)
 & \pog{} (2) & \pens{} (3) & \netflix{} (4) & \books{} (1) & \lastfm{} (5) \\
\midrule
\pog{} 
& \cellcolor[HTML]{d5d5d5} 0.5354 & \cellcolor{bgGood} 0.6216 & \cellcolor{bgGood} 0.7460 (3) & \cellcolor[HTML]{ffffff} 0.6649 & \cellcolor{bgGood} 0.6702 
& \cellcolor[HTML]{d5d5d5} 0.5147 & \cellcolor[HTML]{ffffff} 0.4937 & \cellcolor{bgGood} 0.5192 & \cellcolor{bgGood} 0.5660 (5) & \cellcolor[HTML]{ffffff} 0.4927 
\\
\pens{} 
& \cellcolor[HTML]{ffffff} 0.5890 & \cellcolor[HTML]{d5d5d5} 0.6311 & \cellcolor{bgGood} 0.7486 (2) & \cellcolor{bgGood} 0.6929 & \cellcolor[HTML]{ffffff} 0.5829 
& \cellcolor{bgGood} 0.5295 & \cellcolor[HTML]{d5d5d5} 0.5134 & \cellcolor{bgGood} 0.5186 & \cellcolor{bgGood} 0.5859 (1) & \cellcolor[HTML]{ffffff} 0.5077 
\\
\netflix{} 
& \cellcolor{bgGood} 0.7054 & \cellcolor[HTML]{ffffff} 0.7012 & \cellcolor[HTML]{d5d5d5} 0.7422 (5) & \cellcolor[HTML]{ffffff} 0.7144 & \cellcolor[HTML]{ffffff} 0.7110 
& \cellcolor[HTML]{ffffff} 0.5130 & \cellcolor[HTML]{ffffff} 0.5138 & \cellcolor[HTML]{d5d5d5} 0.4958 & \cellcolor{bgGood} 0.5725 (4) & \cellcolor[HTML]{ffffff} 0.4825 
\\
\books{} 
& \cellcolor[HTML]{ffffff} 0.7061 & \cellcolor[HTML]{ffffff} 0.6832 & \cellcolor{bgGood} 0.7458 (4) & \cellcolor[HTML]{d5d5d5} 0.6935 & \cellcolor{bgGood} 0.6843 
& \cellcolor[HTML]{ffffff} 0.5597 & \cellcolor[HTML]{ffffff} 0.5249 & \cellcolor[HTML]{ffffff} 0.5062 & \cellcolor[HTML]{d5d5d5} 0.5727 (3) & \cellcolor[HTML]{ffffff} 0.5089
\\
\lastfm{} 
& \cellcolor[HTML]{ffffff} 0.6406 & \cellcolor{bgGood} 0.6688 & \cellcolor{bgGood} 0.7487 (1) & \cellcolor[HTML]{ffffff} 0.6737 & \cellcolor[HTML]{d5d5d5} 0.6736 
& \cellcolor{bgGood} 0.5133 & \cellcolor{bgGood} 0.5218 & \cellcolor{bgGood} 0.4950 & \cellcolor{bgGood} 0.5754 (2) & \cellcolor[HTML]{d5d5d5} 0.4818 
\\
\midrule[1pt]
& \multicolumn{5}{c|}{\steam{}} & \multicolumn{5}{c}{\electronics{}} \\
\midrule[1pt]
& \pog{} (5) & \pens{} (2) & \netflix{} (3) & \books{} (4) & \lastfm{} (1)
& \pog{} (5) & \pens{} (4) & \netflix{} (2) & \books{} (1) & \lastfm{} (3) \\
\midrule
\pog{} 
& \cellcolor[HTML]{d5d5d5} 0.5684 & \cellcolor{bgGood} 0.8902 & \cellcolor[HTML]{ffffff} 0.8679 & \cellcolor[HTML]{ffffff} 0.6110 & \cellcolor{bgGood} 0.9299 (3)
& \cellcolor[HTML]{d5d5d5} 0.4896 & \cellcolor[HTML]{ffffff} 0.5255 & \cellcolor{bgGood} 0.5720 & \cellcolor{bgGood} 0.6274 (4) & \cellcolor[HTML]{ffffff} 0.5340 
\\
\pens{} 
& \cellcolor[HTML]{ffffff} 0.8452 & \cellcolor[HTML]{d5d5d5} 0.8979 & \cellcolor[HTML]{ffffff} 0.8781 & \cellcolor[HTML]{ffffff} 0.7137 & \cellcolor[HTML]{ffffff} 0.8827 
& \cellcolor{bgGood} 0.5426 & \cellcolor[HTML]{d5d5d5} 0.5353 & \cellcolor{bgGood} 0.5885 & \cellcolor{bgGood} 0.6380 (2) & \cellcolor[HTML]{ffffff} 0.5211 
\\
\netflix{} 
& \cellcolor{bgGood} 0.8826 & \cellcolor{bgGood} 0.9076 & \cellcolor[HTML]{d5d5d5} 0.8547 & \cellcolor[HTML]{ffffff} 0.7431 & \cellcolor{bgGood} 0.9297 (4)
& \cellcolor[HTML]{ffffff} 0.5559 & \cellcolor[HTML]{ffffff} 0.5468 & \cellcolor[HTML]{d5d5d5} 0.5947 & \cellcolor{bgGood} 0.6419 (1) & \cellcolor[HTML]{ffffff} 0.5462 
\\
\books{} 
& \cellcolor{bgGood} 0.7833 & \cellcolor{bgGood} 0.9043 & \cellcolor{bgGood} 0.8674 & \cellcolor[HTML]{d5d5d5} 0.6373 & \cellcolor{bgGood} 0.9307 (2)
& \cellcolor[HTML]{ffffff} 0.5741 & \cellcolor[HTML]{ffffff} 0.5564 & \cellcolor[HTML]{ffffff} 0.6078 & \cellcolor[HTML]{d5d5d5} 0.6267 (5) & \cellcolor[HTML]{ffffff} 0.5506
\\
\lastfm{} 
& \cellcolor[HTML]{ffffff} 0.8633 & \cellcolor{bgGood} 0.9191 (5) & \cellcolor[HTML]{ffffff} 0.8899 & \cellcolor[HTML]{ffffff} 0.6743 & \cellcolor[HTML]{d5d5d5} 0.9363 (1)
& \cellcolor{bgGood} 0.5445 & \cellcolor{bgGood} 0.5290 & \cellcolor{bgGood} 0.5918 & \cellcolor{bgGood} 0.6294 (3) & \cellcolor[HTML]{d5d5d5} 0.5414 
\\
\midrule[1pt]
& \multicolumn{5}{c|}{\hotelrec{}} & \multicolumn{5}{c}{Overall} \\
\midrule[1pt]
& \pog{} (3) & \pens{} (5) & \netflix{} (2) & \books{} (1) & \lastfm{} (4)
& \pog{} (5) & \pens{} (4) & \netflix{} (1) & \books{} (3) & \lastfm{} (2) \\
\midrule
\pog{} 
& \cellcolor[HTML]{d5d5d5} 0.5142 & \cellcolor[HTML]{ffffff} 0.4903 & \cellcolor{bgGood} 0.5947 & \cellcolor{bgGood} 0.6004 & \cellcolor{bgGood} 0.4990 
& \cellcolor[HTML]{d5d5d5} 0.5245 & \cellcolor[HTML]{ffffff} 0.6043 & \cellcolor{bgGood} 0.6600 & \cellcolor[HTML]{ffffff} 0.6139 & \cellcolor{bgGood} 0.6252
\\
\pens{} 
& \cellcolor{bgGood} 0.5006 & \cellcolor[HTML]{d5d5d5} 0.4860 & \cellcolor{bgGood} 0.5892 & \cellcolor{bgGood} 0.6156 (2) & \cellcolor{bgGood} 0.5040 
& \cellcolor{bgGood} 0.6092 & \cellcolor[HTML]{d5d5d5} 0.6127 & \cellcolor{bgGood} 0.6646 (2) & \cellcolor{bgGood} 0.6492 & \cellcolor[HTML]{ffffff} 0.5997 
\\
\netflix{} 
& \cellcolor[HTML]{ffffff} 0.5399 & \cellcolor[HTML]{ffffff} 0.5188 & \cellcolor[HTML]{d5d5d5} 0.6104 (4) & \cellcolor{bgGood} 0.6204 (1) & \cellcolor[HTML]{ffffff} 0.5189 
& \cellcolor[HTML]{ffffff} 0.6394 & \cellcolor[HTML]{ffffff} 0.6376 & \cellcolor[HTML]{d5d5d5} 0.6596 (4) & \cellcolor[HTML]{ffffff} 0.6585 & \cellcolor[HTML]{ffffff} 0.6377 
\\
\books{} 
& \cellcolor[HTML]{ffffff} 0.5622 & \cellcolor[HTML]{ffffff} 0.5033 & \cellcolor[HTML]{ffffff} 0.6006 & \cellcolor[HTML]{d5d5d5} 0.6149 (3) & \cellcolor[HTML]{ffffff} 0.5216 
& \cellcolor{bgGood} 0.6371 & \cellcolor[HTML]{ffffff} 0.6344 & \cellcolor{bgGood} 0.6656 (1) & \cellcolor[HTML]{d5d5d5} 0.6290 & \cellcolor{bgGood} 0.6392 (5)
\\
\lastfm{} 
& \cellcolor[HTML]{ffffff} 0.4921 & \cellcolor[HTML]{ffffff} 0.4875 & \cellcolor{bgGood} 0.5903 & \cellcolor{bgGood} 0.6055 (5) & \cellcolor[HTML]{d5d5d5} 0.5125 
& 0.6108 & \cellcolor{bgGood} 0.6252 & \cellcolor{bgGood} 0.6631 (3) & \cellcolor[HTML]{ffffff} 0.6317 & \cellcolor[HTML]{d5d5d5} 0.6291 
\\
\midrule
\bottomrule
\end{tabular}
}
\end{table}

\begin{figure}
    \centering
    
\begin{tikzpicture}
\begin{axis}[
    ybar=0pt,
    bar width=10pt,
    width=12cm,
    height=7cm,
    ymin=0,
    ymax=40,
    enlarge x limits=0.15,
    ylabel={Top-5 Rank Count},
    symbolic x coords={Rank-1, Rank-2, Rank-3, Rank-4, Rank-5},
    xtick=data,
    major grid style={gray!20},
    grid=major,
    legend style={at={(0.5,-0.15)}, anchor=north, legend columns=-1},
    axis background/.style={fill=gray!5},
    nodes near coords,
    every node near coord/.append style={font=\footnotesize},
    tick label style={font=\small},
]

\addplot+[style={firstordercolor, fill=firstordercolor}] 
    coordinates {(Rank-1,4) (Rank-2,9) (Rank-3,10) (Rank-4,10) (Rank-5,7)};

\addplot+[style={secondordercolor, fill=secondordercolor}] 
    coordinates {(Rank-1,33) (Rank-2,4) (Rank-3,1) (Rank-4,1) (Rank-5,1)};

\legend{First Step, Second Step}
\end{axis}
\end{tikzpicture}
    
    \caption{Impact of fine-tuning dataset order in multi-domain fine-tuning. The statistics are derived from Table~\ref{tab:order}. For each test set, we select the top-5 AUC scores from the 25 multi-domain combinations (excluding single-dataset fine-tuning results). For each top-5 result (e.g., 0.6687 on \mind{}), where fine-tuning is performed first on dataset $x$ (e.g., \pog{}) and then on $y$ (e.g., \pens{}), we record the single-dataset performance ranks of $x$ and $y$ (e.g., Rank-4 for \pog{}, Rank-1 for \pens{}). We then increment the corresponding counts in the First Step and Second Step categories. For example, a count of 33 for Rank-1 in the Second Step indicates that, across 10 test sets and considering the top 5 entries from each set, the best-performing dataset from single-dataset fine-tuning was placed in the second step 33 times.
    }
    \label{fig:order}
\end{figure}
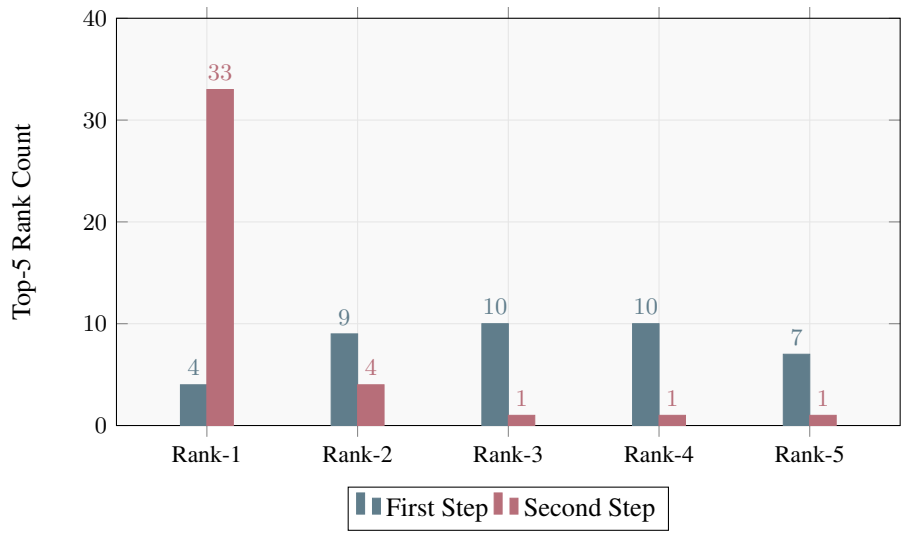

\begin{figure}
    \centering
    
\begin{tikzpicture}
\begin{axis}[
    ybar,
    bar width=12pt,
    width=14cm,
    height=7cm,
    ymin=0,
    ymax=10,
    enlarge x limits=0.1,
    ylabel={Top-5 Rank Count},
    xlabel={Fine-tune Pairs},
    symbolic x coords={2-1, 4-1, 3-1, 5-1, 3-2, 4-2, 1-2, 1-3, 1-4, 1-5},
    xtick=data,
    x tick label style={rotate=45, anchor=east, font=\small},
    major grid style={gray!20},
    grid=major,
    axis background/.style={fill=gray!5},
    nodes near coords,
    every node near coord/.append style={font=\footnotesize},
    tick label style={font=\small},
]

\addplot+[style={barcolor, fill=barcolor}] 
    coordinates {
        (2-1,9) (4-1,9) (3-1,8) (5-1,7)
        (3-2,2) (4-2,1) (1-3,1) (1-4,1) (1-5,1) (1-2,1)
    };

\end{axis}
\end{tikzpicture}
    
    \caption{Impact of fine-tuning dataset order in multi-domain fine-tuning. The statistics are derived from Table~\ref{tab:order}. For each test set, we select the top-5 AUC scores from the 25 multi-domain combinations (excluding single-dataset baselines). For each top-5 result (e.g., 0.6687 on \mind{}), where fine-tuning is performed first on dataset $x$ (e.g., \pog{}) and then on $y$ (e.g., \pens{}), we record the single-dataset performance ranks of $x$ and $y$ (e.g., Rank-4 for \pog{}, Rank-1 for \pens{}). We then increment the corresponding counts for such combination. For example, the count of 9 in the 4-1 pair means that, out of all top-5 results across 10 test sets, there were 9 cases where the fourth-ranked dataset (in single-dataset fine-tuning) was used first, followed by the best-performing dataset second.}
    \label{fig:pairs}
\end{figure}
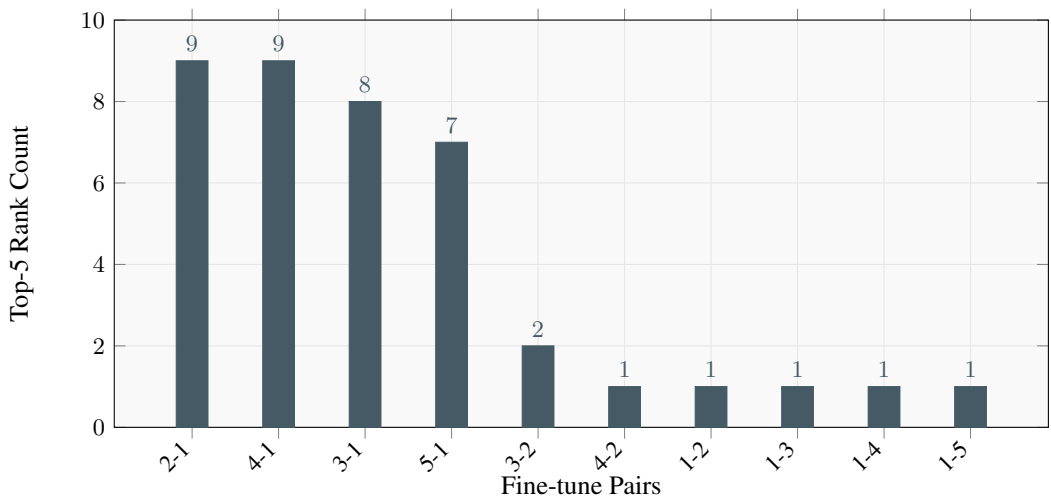

\subsection{Impact of Fine-tuning Dataset Order In Multi-domain Recommendation}

Previously, Table~\ref{tab:multi-domain} compares three evaluation strategies: \rB{}, \rG{}, and \rH{}. Both \rG{} and \rH{} involve training on a mixture of all available fine-tune sets. In contrast, we analyze a sequential fine-tuning strategy here, focusing on how to select datasets and determine fine-tuning order. Given the combinatorial complexity of all five datasets ($A_5^5$), we restrict our analysis to pairs of fine-tune sets. Based on the results from Table~\ref{tab:order}, we can observe that:

\textbf{First}, in most cases, the value at row $i$, column $j$ exceeds that of the corresponding single-dataset fine-tuning result in row $j$, column $j$, indicating that using two datasets generally provides greater benefit than using only one. However, it does not necessarily surpass the value at row $i$, column $i$, since knowledge learned from dataset $i$, the first step, may be subject to catastrophic forgetting during continual fine-tuning with dataset $j$.

\textbf{Second}, building on this observation, the dataset used in the later stage (second step) of fine-tuning tends to have a dominant influence on the final performance. For example, in columns corresponding to datasets that achieved the best single-dataset results, green highlights are commonly observed (e.g., the \pens{} column when the test set is \hm{}, or the \books{} column when the test set is \goodreads{}). To further investigate this effect, we present a more detailed analysis in Figure~\ref{fig:order}, showing that datasets with stronger single-dataset performance are generally more effective when used in the second fine-tuning step.

\textbf{Third}, we further investigate which fine-tuning combinations are most likely to yield Top-5 performance. We hypothesize that this is related to the single-dataset performance of the fine-tune sets. To examine this, we present Figure~\ref{fig:pairs}. The results suggest that using a lower-ranked dataset in the first step, followed by the top-performing dataset in the second step, tends to produce the best outcomes for the target test set.

\begin{table}[]
\centering
\caption{Performance comparison in multi-domain recommendation scenario, with the evaluation metrics. Experiments are conducted on the \mind{} and \microlens{} datasets. We bold the best results for each metric.}\label{tab:metric-1}

\resizebox{.9\linewidth}{!}{
\begin{tabular}{ll|llllll}
\toprule
First Step     & Second Step    & AUC    & nDCG@1 & nDCG@5 & MRR    & Recall@1 & Recall@5 \\
\midrule[1pt]
\multicolumn{8}{c}{\mind{}} \\
\midrule[1pt]
\pog{}     & \pens{}    & 0.6687 & 0.5179 & 0.5767 & 0.5687 & 0.1865   & 0.5808   \\
\pens{}    & \pog{}     & 0.6636 & 0.4905 & 0.5653 & 0.5513 & 0.1653   & 0.5643   \\
\midrule
\pog{}     & \netflix{} & 0.5221 & 0.3234 & 0.4072 & 0.4333 & 0.1100   & 0.4249   \\
\netflix{} & \pog{}     & 0.5530 & 0.3461 & 0.4401 & 0.4607 & 0.1232   & 0.4711   \\
\midrule
\pog{}     & \books{}   & 0.4863 & 0.2675 & 0.3636 & 0.4009 & 0.0866   & 0.3944   \\
\books{}   & \pog{}     & 0.5248 & 0.3154 & 0.4090 & 0.4330 & 0.1001   & 0.4365   \\
\midrule
\pog{}     & \lastfm{}  & 0.5901 & 0.4206 & 0.4918 & 0.5019 & 0.1514   & 0.5071   \\
\lastfm{}  & \pog{}     & 0.5440 & 0.3506 & 0.4377 & 0.4562 & 0.1170   & 0.4644   \\
\midrule
\pens{}    & \books{}   & 0.5270 & 0.2999 & 0.3987 & 0.4309 & 0.1042   & 0.4360   \\
\books{}   & \pens{}    & \textbf{0.6759} & \textbf{0.5342} & 0.5811 & \textbf{0.5752} & \textbf{0.1903}   & 0.5777   \\
\midrule
\pens{}    & \netflix{} & 0.6186 & 0.4205 & 0.5067 & 0.5110 & 0.1451   & 0.5228   \\
\netflix{} & \pens{}    & 0.6757 & 0.5208 & \textbf{0.5813} & 0.5743 & 0.1847   & \textbf{0.5886}   \\
\midrule
\pens{}    & \lastfm{}  & 0.6482 & 0.4630 & 0.5440 & 0.5319 & 0.1539   & 0.5597   \\
\lastfm{}  & \pens{}    & 0.6738 & 0.5182 & 0.5777 & 0.5719 & 0.1813   & 0.5814   \\
\midrule
\books{}   & \netflix{} & 0.5223 & 0.2990 & 0.4026 & 0.4289 & 0.0973   & 0.4326   \\
\netflix{} & \books{}   & 0.4992 & 0.2688 & 0.3774 & 0.4150 & 0.0904   & 0.4128   \\
\midrule
\books{}   & \lastfm{}  & 0.5785 & 0.3932 & 0.4759 & 0.4878 & 0.1373   & 0.5025   \\
\lastfm{}  & \books{}   & 0.5149 & 0.3005 & 0.3941 & 0.4294 & 0.1057   & 0.4276   \\
\midrule
\netflix{} & \lastfm{}  & 0.5866 & 0.4026 & 0.4821 & 0.4958 & 0.1428   & 0.5032   \\
\lastfm{}  & \netflix{} & 0.5039 & 0.3012 & 0.3857 & 0.4249 & 0.1013   & 0.4178  \\
\midrule[1pt]
\multicolumn{8}{c}{\microlens{}} \\
\midrule[1pt]
\pog{}     & \pens{}    & 0.6443 & 0.6782 & 0.6782 & 0.7648 & 0.3301   & 1.0000   \\
\pens{}    & \pog{}     & 0.6262 & 0.6638 & 0.8450 & 0.7433 & 0.3107   & 1.0000   \\
\midrule
\pog{}     & \netflix{} & 0.5753 & 0.5945 & 0.8201 & 0.7204 & 0.2873   & 1.0000   \\
\netflix{} & \pog{}     & 0.6222 & 0.6665 & 0.8441 & 0.7554 & 0.3292   & 1.0000   \\
\midrule
\pog{}     & \books{}   & 0.5304 & 0.5257 & 0.7972 & 0.6889 & 0.2511   & 1.0000   \\
\books{}   & \pog{}     & 0.6328 & 0.6635 & 0.8463 & 0.7607 & 0.3276   & 1.0000   \\
\midrule
\pog{}     & \lastfm{}  & 0.6495 & 0.6867 & 0.6867 & 0.7749 & 0.3423   & 1.0000   \\
\lastfm{}  & \pog{}     & 0.6363 & 0.6705 & 0.8484 & 0.7622 & 0.3307   & 1.0000   \\
\midrule
\pens{}    & \books{}   & 0.5508 & 0.5653 & 0.8096 & 0.7094 & 0.2780   & 1.0000   \\
\books{}   & \pens{}    & 0.6662 & 0.7094 & 0.8628 & 0.7828 & 0.3496   & 1.0000   \\
\midrule
\pens{}    & \netflix{} & 0.6157 & 0.6423 & 0.8387 & 0.7473 & 0.3137   & 1.0000   \\
\netflix{} & \pens{}    & 0.6693 & 0.7028 & 0.8624 & 0.7850 & 0.3483   & 1.0000   \\
\midrule
\pens{}    & \lastfm{}  & 0.6439 & 0.6742 & 0.8511 & 0.7619 & 0.3258   & 1.0000   \\
\lastfm{}  & \pens{}    & \textbf{0.6779} & \textbf{0.7162} & \textbf{0.8671} & \textbf{0.7921} & \textbf{0.3550}   & 1.0000   \\
\midrule
\books{}   & \netflix{} & 0.5784 & 0.5939 & 0.8211 & 0.7248 & 0.2913   & 1.0000   \\
\netflix{} & \books{}   & 0.5650 & 0.5841 & 0.8161 & 0.7192 & 0.2868   & 1.0000   \\
\midrule
\books{}   & \lastfm{}  & 0.6462 & 0.6879 & 0.8540 & 0.7722 & 0.3413   & 1.0000   \\
\lastfm{}  & \books{}   & 0.5338 & 0.5435 & 0.8011 & 0.6999 & 0.2687   & 1.0000   \\
\midrule
\netflix{} & \lastfm{}  & 0.6688 & 0.7104 & 0.8636 & 0.7873 & 0.3534   & 1.0000   \\
\lastfm{}  & \netflix{} & 0.5875 & 0.6080 & 0.8251 & 0.7334 & 0.3019   & 1.0000   \\
\bottomrule

\end{tabular}
}
\end{table}

\begin{table}[]
\centering
\caption{Performance comparison in multi-domain recommendation scenario, with the evaluation metrics. Experiments are conducted on the \microlens{} and \yelp{} datasets. We bold the best results for each metric.}\label{tab:metric-2}

\resizebox{.9\linewidth}{!}{
\begin{tabular}{ll|llllll}
\toprule
First Step     & Second Step    & AUC    & nDCG@1 & nDCG@5 & MRR    & Recall@1 & Recall@5 \\
\midrule[1pt]
\multicolumn{8}{c}{\movielens{}} \\
\midrule[1pt]
\pog{}     & \pens{}    & 0.6216 & 0.4486 & 0.5870 & 0.5467 & 0.2146   & 0.7081   \\
\pens{}    & \pog{}     & 0.5890 & 0.3970 & 0.5600 & 0.5149 & 0.1873   & 0.6875   \\
\midrule
\pog{}     & \netflix{} & 0.7460 & 0.5643 & 0.7022 & 0.6516 & 0.2757   & 0.8207   \\
\netflix{} & \pog{}     & 0.7054 & 0.5360 & 0.6671 & 0.6224 & 0.2648   & 0.7896   \\
\midrule
\pog{}     & \books{}   & 0.6649 & 0.4753 & 0.6223 & 0.5757 & 0.2297   & 0.7523   \\
\books{}   & \pog{}     & 0.7061 & 0.5277 & 0.6685 & 0.6181 & 0.2592   & 0.7952   \\
\midrule
\pog{}     & \lastfm{}  & 0.6702 & 0.5270 & 0.6434 & 0.6072 & 0.2610   & 0.7577   \\
\lastfm{}  & \pog{}     & 0.6406 & 0.4749 & 0.6085 & 0.5677 & 0.2337   & 0.7323   \\
\midrule
\pens{}    & \books{}   & 0.6929 & 0.5158 & 0.6487 & 0.6072 & 0.2552   & 0.7728   \\
\books{}   & \pens{}    & 0.6832 & 0.5330 & 0.6531 & 0.6079 & 0.2594   & 0.7668   \\
\midrule
\pens{}    & \netflix{} & 0.7486 & 0.5926 & 0.7075 & 0.6648 & 0.2934   & 0.8187   \\
\netflix{} & \pens{}    & 0.7012 & 0.5546 & 0.6680 & 0.6262 & 0.2732   & 0.7790   \\
\midrule
\pens{}    & \lastfm{}  & 0.5829 & 0.3782 & 0.5432 & 0.4812 & 0.1570   & 0.6479   \\
\lastfm{}  & \pens{}    & 0.6688 & 0.5217 & 0.6417 & 0.6024 & 0.2559   & 0.7530   \\
\midrule
\books{}   & \netflix{} & 0.7458 & 0.5815 & 0.7041 & 0.6576 & 0.2861   & 0.8183   \\
\netflix{} & \books{}   & 0.7144 & 0.5544 & 0.6732 & 0.6279 & 0.2723   & 0.7901   \\
\midrule
\books{}   & \lastfm{}  & 0.6843 & 0.5417 & 0.6553 & 0.6196 & 0.2690   & 0.7688   \\
\lastfm{}  & \books{}   & 0.6737 & 0.4782 & 0.6307 & 0.5871 & 0.2357   & 0.7659   \\
\midrule
\netflix{} & \lastfm{}  & 0.7110 & 0.5660 & 0.6785 & 0.6417 & 0.2823   & 0.7916   \\
\lastfm{}  & \netflix{} & \textbf{0.7487} & \textbf{0.5902} & \textbf{0.7092} & \textbf{0.6657} & \textbf{0.2934}   & \textbf{0.8239}  \\
\midrule[1pt]
\multicolumn{8}{c}{\yelp{}} \\
\midrule[1pt]
\pog{}     & \pens{}    & 0.4937 & 0.4761 & 0.7172 & 0.6378 & 0.2227   & 0.8871   \\
\pens{}    & \pog{}     & 0.5295 & 0.5086 & 0.7383 & 0.6575 & 0.2385   & 0.9011   \\
\midrule
\pog{}     & \netflix{} & 0.5192 & 0.5164 & 0.7356 & 0.6576 & 0.2461   & 0.8985   \\
\netflix{} & \pog{}     & 0.5130 & 0.4964 & 0.7289 & 0.6572 & 0.2445   & 0.8998   \\
\midrule
\pog{}     & \books{}   & 0.5660 & 0.5535 & 0.7599 & 0.6842 & 0.2684   & 0.9166   \\
\books{}   & \pog{}     & 0.5597 & 0.5421 & 0.7553 & 0.6813 & 0.2645   & 0.9152   \\
\midrule
\pog{}     & \lastfm{}  & 0.4927 & 0.4765 & 0.7168 & 0.6465 & 0.2360   & 0.8917   \\
\lastfm{}  & \pog{}     & 0.5133 & 0.5043 & 0.7281 & 0.6559 & 0.2451   & 0.8938   \\
\midrule
\pens{}    & \books{}   & 0.5859 & 0.5732 & 0.7695 & 0.6981 & 0.2815   & 0.9208   \\
\books{}   & \pens{}    & 0.5249 & 0.5097 & 0.7358 & 0.6605 & 0.2461   & 0.9022   \\
\midrule
\pens{}    & \netflix{} & 0.5186 & 0.5078 & 0.7340 & 0.6587 & 0.2476   & 0.9009   \\
\netflix{} & \pens{}    & 0.5138 & 0.5029 & 0.7303 & 0.6552 & 0.2437   & 0.9002   \\
\midrule
\pens{}    & \lastfm{}  & 0.5077 & 0.4995 & 0.7262 & 0.6482 & 0.2365   & 0.8930   \\
\lastfm{}  & \pens{}    & 0.5218 & 0.5019 & 0.7321 & 0.6595 & 0.2441   & 0.9003   \\
\midrule
\books{}   & \netflix{} & 0.5062 & 0.5105 & 0.7294 & 0.6543 & 0.2467   & 0.8953   \\
\netflix{} & \books{}   & 0.5725 & \textbf{0.5635} & 0.7640 & 0.6909 & 0.2752   & 0.9175   \\
\midrule
\books{}   & \lastfm{}  & 0.5089 & 0.4963 & 0.7266 & 0.6568 & 0.2453   & 0.8972   \\
\lastfm{}  & \books{}   & \textbf{0.5754} & 0.5628 & \textbf{0.7657} & \textbf{0.6939} & \textbf{0.2770}   & \textbf{0.9220}   \\
\midrule
\netflix{} & \lastfm{}  & 0.4825 & 0.4684 & 0.7091 & 0.6412 & 0.2320   & 0.8850   \\
\lastfm{}  & \netflix{} & 0.4950 & 0.4885 & 0.7211 & 0.6491 & 0.2412   & 0.8955  \\
\bottomrule
\end{tabular}
}
\end{table}

\begin{table}[]
\centering
\caption{Performance comparison in multi-domain recommendation scenario, with the evaluation metrics. Experiments are conducted on the \goodreads{} and \cds{} datasets. We bold the best results for each metric.}\label{tab:metric-3}

\resizebox{.9\linewidth}{!}{
\begin{tabular}{ll|llllll}
\toprule
First Step     & Second Step    & AUC    & nDCG@1 & nDCG@5 & MRR    & Recall@1 & Recall@5 \\
\midrule[1pt]
\multicolumn{8}{c}{\goodreads{}} \\
\midrule[1pt]
\pog{}     & \pens{}    & 0.5659 & 0.2801 & 0.4068 & 0.4011 & 0.1319   & 0.5098   \\
\pens{}    & \pog{}     & 0.5632 & 0.2638 & 0.4017 & 0.3902 & 0.1224   & 0.5006   \\
\midrule
\pog{}     & \netflix{} & 0.5631 & 0.2773 & 0.4025 & 0.3992 & 0.1322   & 0.5060   \\
\netflix{} & \pog{}     & 0.5690 & 0.2705 & 0.4091 & 0.4026 & 0.1316   & 0.5236   \\
\midrule
\pog{}     & \books{}   & 0.5700 & 0.2913 & 0.4137 & 0.4081 & 0.1391   & 0.5173   \\
\books{}   & \pog{}     & 0.5822 & 0.3072 & 0.4315 & 0.4214 & 0.1483   & 0.5418   \\
\midrule
\pog{}     & \lastfm{}  & 0.5858 & 0.2866 & 0.4311 & 0.4247 & 0.1420   & 0.5541   \\
\lastfm{}  & \pog{}     & 0.5540 & 0.2752 & 0.3944 & 0.3972 & 0.1339   & 0.4980   \\
\midrule
\pens{}    & \books{}   & 0.6024 & 0.3263 & 0.4463 & 0.4380 & 0.1604   & 0.5559   \\
\books{}   & \pens{}    & 0.5947 & 0.3389 & 0.4488 & 0.4387 & 0.1619   & 0.5490   \\
\midrule
\pens{}    & \netflix{} & 0.5810 & 0.3013 & 0.4289 & 0.4209 & 0.1475   & 0.5383   \\
\netflix{} & \pens{}    & 0.5721 & 0.2969 & 0.4172 & 0.4155 & 0.1434   & 0.5228   \\
\midrule
\pens{}    & \lastfm{}  & 0.5375 & 0.2373 & 0.3741 & 0.3605 & 0.0994   & 0.4657   \\
\lastfm{}  & \pens{}    & 0.5728 & 0.2925 & 0.4182 & 0.4143 & 0.1408   & 0.5236   \\
\midrule
\books{}   & \netflix{} & 0.5950 & 0.3267 & 0.4455 & 0.4354 & 0.1596   & 0.5504   \\
\netflix{} & \books{}   & \textbf{0.5973} & \textbf{0.3559} & \textbf{0.4529} & \textbf{0.4440} & \textbf{0.1743}   & 0.5487   \\
\midrule
\books{}   & \lastfm{}  & 0.5864 & 0.3162 & 0.4433 & 0.4342 & 0.1567   & 0.5616   \\
\lastfm{}  & \books{}   & 0.5953 & 0.3181 & 0.4452 & 0.4346 & 0.1552   & \textbf{0.5588}   \\
\midrule
\netflix{} & \lastfm{}  & 0.5743 & 0.2949 & 0.4227 & 0.4212 & 0.1463   & 0.5397   \\
\lastfm{}  & \netflix{} & 0.5748 & 0.3099 & 0.4234 & 0.4224 & 0.1526   & 0.5294  \\
\midrule[1pt]
\multicolumn{8}{c}{\cds{}} \\
\midrule[1pt]
\pog{}     & \pens{}    & 0.5225 & 0.5763 & 0.7969 & 0.7180 & 0.2760   & 0.9523   \\
\pens{}    & \pog{}     & 0.5140 & 0.5798 & 0.7964 & 0.7105 & 0.2706   & 0.9515   \\
\midrule
\pog{}     & \netflix{} & 0.5254 & 0.5911 & 0.8011 & 0.7250 & 0.2865   & 0.9537   \\
\netflix{} & \pog{}     & 0.5237 & 0.5805 & 0.7978 & 0.7238 & 0.2861   & 0.9544   \\
\midrule
\pog{}     & \books{}   & 0.5517 & 0.6209 & 0.8137 & 0.7394 & 0.3027   & 0.9584   \\
\books{}   & \pog{}     & 0.5563 & 0.6182 & 0.8126 & 0.7418 & 0.3030   & 0.9557   \\
\midrule
\pog{}     & \lastfm{}  & 0.4947 & 0.5583 & 0.7864 & 0.7129 & 0.2763   & 0.9495   \\
\lastfm{}  & \pog{}     & 0.5098 & 0.5785 & 0.7941 & 0.7184 & 0.2840   & 0.9515   \\
\midrule
\pens{}    & \books{}   & 0.5551 & 0.6242 & 0.8156 & 0.7443 & 0.3078   & 0.9591   \\
\books{}   & \pens{}    & 0.5340 & 0.5878 & 0.8029 & 0.7276 & 0.2854   & 0.9553   \\
\midrule
\pens{}    & \netflix{} & 0.5366 & 0.5969 & 0.8038 & 0.7326 & 0.2941   & 0.9534   \\
\netflix{} & \pens{}    & 0.5293 & 0.5829 & 0.8001 & 0.7259 & 0.2845   & 0.9548   \\
\midrule
\pens{}    & \lastfm{}  & 0.5023 & 0.5619 & 0.7910 & 0.7171 & 0.2756   & 0.9511   \\
\lastfm{}  & \pens{}    & 0.5263 & 0.5866 & 0.7974 & 0.7258 & 0.2879   & 0.9497   \\
\midrule
\books{}   & \netflix{} & 0.5387 & 0.6029 & 0.8064 & 0.7337 & 0.2948   & 0.9555   \\
\netflix{} & \books{}   & 0.5602 & \textbf{0.6296} & 0.8170 & 0.7453 & 0.3100   & 0.9590   \\
\midrule
\books{}   & \lastfm{}  & 0.5039 & 0.5656 & 0.7886 & 0.7174 & 0.2809   & 0.9477   \\
\lastfm{}  & \books{}   & \textbf{0.5654} & 0.6278 & \textbf{0.8190} & \textbf{0.7492} & \textbf{0.3114}   & \textbf{0.9625}   \\
\midrule
\netflix{} & \lastfm{}  & 0.5098 & 0.5785 & 0.7941 & 0.7184 & 0.2840   & 0.9515   \\
\lastfm{}  & \netflix{} & 0.5259 & 0.5927 & 0.8001 & 0.7298 & 0.2942   & 0.9522   \\
\bottomrule
\end{tabular}
}
\end{table}

\subsection{Additional Evaluation Metrics}

In the main text, we report only the AUC metric due to the space constraints. Here, we provide additional evaluation metrics, including nDCG@1, nDCG@5, MRR, Recall@1, and Recall@5, for a more comprehensive comparison.

As shown in Table~\ref{tab:metric-1}, Table~\ref{tab:metric-2}, and Table~\ref{tab:metric-3}, other metrics generally align with the AUC results, supporting the consistency of our findings. We will release the complete experimental results on our website.

\end{document}